%% file: doctorat_arxiv.tex
\documentclass[a4paper,12pt,oneside]{book}

\usepackage{bioinformatics}
\usepackage{doublespace}
\usepackage{graphicx}
\begin{document}
\begin{singlespace}
\begin{titlepage}
\begin{center}
\LARGE
{\bf
``Sequence-Structure Relationship in Proteins: a Computational Analysis of Proteins
that Differ in Sequence but Share the Same Fold''
\\[8cm]
\Large
Thesis submitted for the Degree \\
Doctor of Philosophy \\
by Iddo Friedberg
\\[5cm]
\normalsize
Submitted to the Senate of the Hebrew University \\
July 2002
}
\end{center}
\end{titlepage}
\renewcommand{\thepage}{\roman{page}}
This work was carried out under the supervision of Prof. Hanah Margalit.
\newpage
\Large \underline{\bf Acknowledgements}
\normalsize
\\[2cm]
\input{ack.tex}

\end{singlespace}
\renewcommand{\thepage}{\roman{page}}
\chapter*{Summary}
\label{chap:summary}
\markright{SUMMARY}
\input{summary_arxiv.tex}
\tableofcontents
\chapter{Introduction}
\label{chap:introduction}
\renewcommand{\thepage}{\arabic{page}}
\setcounter{page}{1}

How does a protein sequence determine its structure? This question is known
as the protein folding problem. Currently, this is an open problem in
structural biology.  The underlying assumption of the protein folding
problem is that the native folded state is encoded in the amino-acid
sequence. The goal is to decode this information, so that given a protein
sequence, its structure may be made known.

The dogma stating that the amino-acid sequence determines the native folding
state was formulated by Christian Anfinsen
\cite{Anfinsen_etal:62,Anfinsen:73}.  It is based on studies in which
Anfinsen \textit{et al.} have shown that denatured RNaseA returns
spontaneously to a fully functional form. Since then, this phenomenon was
exhibited in many other proteins. It is known that proteins may be
irreversibly denatured under severe conditions, or my not be able to achieve
their native folding state without some kinetic assistance (e.g.
chaperone-mediated folding). There are also numerous examples of protein
misfolding, in which a protein folds into a conformation other than its
native state, or has ``alternative native states''. The best-known cases are
those in which protein misfolding is involved in diseases, e.g.  Alzheimer's
disease and prion diseases. (Reviewed in Cohen, 1999).  These cases,
however, are not contradictory to the sequence determines structure dogma.
In the case of chaperone-mediated folding, the information for achieving a
global free energy minimum, exists in the sequence.  Chaperones merely
assist the folding process, which cannot take place spontaneously due to
kinetic constraints. As such, they act as catalysts, and do not violate the
central dogma of folding \cite{Branden_Tooze:98}.  The ``sequence determines
structure dogma'' still holds also in the other examples, although cases of
discrete alternative conformations for a single sequence or subsequence do
exist \cite{MinorDL_Kim:96,Mezei:98}.

An interesting observation associated with the protein folding problem is
the existence of proteins having a similar structure, but with completely
different amino-acid sequences. This phenomenon raises the following
question: given two or more proteins with dissimilar sequences, but with
similar folds, which positions along the sequence enable the preservation of
the fold, and frequently, of function? In the course of my PhD studies, I
have chosen to use bioinformatics methods to answer this question. The
detection and characterization of those positions can contribute to the sum
body of knowledge regarding the role of the sequence in determining
structure.

This chapter elaborates upon the question set in the previous paragraph, and
places it within the context of the current body of knowledge. It begins
with an overview of current data which exists regarding protein sequences
and structures. This is followed by a review of experimental and
computational studies for the detection of critical amino-acid positions.
Following that, the methods used to carry out this work are described.
Finally, a synopsis of the following chapters, composed of published
research papers is offered.

A technical note: Bioinformatics, being a discipline of its own, carries with
it a unique lexicon. Being a relatively new discipline, some of these terms
suffer from ambiguity. In order to overcome both problems of novelty and
ambiguity, this work has those terms marked in \textbf{boldface} when first
introduced.


\section{Many sequences, few folds}
\label{sec:lots_seq_few_struct}

In many cases, proteins of similar fold and function retain a significant
sequence similarity. Common sequence alignment algorithms enable us to align
and determine the evolutionary relationship between two or more proteins. A
commonly used rule-of-thumb holds that for proteins over 80 residues in
length, a 30\% sequence identity (after alignment) is sufficient to infer a
common fold. However, with the increase of sequence and structural data, it
has become apparent that many proteins share the same fold, and possibly
function, but do not display a detectable sequence similarity. When the
structures of these proteins are examined, their structural similarity is
obvious. Furthermore, in many cases their functional similarity is also
apparent. 

Awareness that structurally similar, sequence dissimilar proteins fold as a common,
actually predominant, phenomenon has only been established recently \cite{Rost:97}.
In that study, Burkhard Rost performed an alignment of all structural homologues
from a database of structural alignments (FSSP, see section \ref{sec:struct_hom}).
He observed one peak at 8.5\% $\pm$ 5\% SD sequence identity. This was quite close
to the random identity, established at 5.6\% $\pm$ 3\% SD. Furthermore, in four
genomes, most close structural homologues have been determined to have less than
45\% pairwise sequence identity. Two conclusions from Rost's study relevant to this
work are summarized here:

\begin{enumerate}

\item Many pairs of similar structures in the PDB have a sequence identity
as low as expected from randomly related sequences. Most structural homologous have
less than 45\% pairwise sequence identity.

\item About 3-4\% of the residues are crucial for protein structure and
function. This was estimated by subtracting the mode of the random
alignments from the mode of structural homologue alignments.

\end{enumerate}

In another study it was shown that 196 domains released in 1998 to the PDB bore no
sequence similarity to existing PDB sequences\cite{Koppensteiner.Lackner.ea:00}.
However, 75\% of the domains were shown to have structural similarity to previously
known folds, and in two-thirds of the cases then similarity in structure coincided
with related function. Brenner and Levitt have shown similar results in their
analysis of new domains incorporated into the PDB over a period of 10 years
\cite{Brenner.Levitt:00}.

\subsection{Sequence family population in protein folds}

A {\bf sequence family} is a collection of sequences which share a sequence
similarity. This is a collection of homologous proteins, sharing a
related structure and function. Sequence families are discussed in Section
\ref{sec:seq_hom}. Different sequence families are assumed to be disjoint.
How many sequence families populate a single fold? And how are they
distributed among folds?

Most folds are populated by a single sequence family.  However, there is a large
number of folds which are populated by more than one sequence family.  In one study
\cite{Wolf_etal:00}, the total number of folds populated by a single sequence
family was estimated to be between 138 and 211. The total number of folds which are
populated by more than one family was estimated to be between 176 and 226. (This
was in 1999, when the total number of known folds was between 331 and 336,
depending on the criteria used for fold classification). Thus, 52-56\% of the
protein folds are populated by more than a single sequence family. Some 12\% of all
folds were estimated to contain more than ten sequence families. Another study has
placed the percentage of folds containing more than one family at 39\%, with
\~{}10\% of all folds estimated to have more than eight sequence families
\cite{Zhang_DeLisi:98}.

The distribution of sequence families in folds has been shown to be
logarithmic, or near-logarithmic \cite{Wolf_etal:00,Zhang_DeLisi:98}. Thus,
while the one family-one fold phenomenon is predominant, still many are
populated by multiple sequence families.

In summary, many folds are populated by dissimilar sequences. It
is not uncommon for proteins sharing only 10\% sequence identity to assume a
similar fold, and often related or identical functions. This suggests that
many positions have no critical role in structure and function
determination, and that folding determinants are restricted only to a
certain small number of residues. Section \ref{sec:history} reviews the
studies, experimental and computational, which were conducted in order to
locate those positions.

\section{Previous studies}
\label{sec:history}

The actual location of critical residues for fold and function can be found
by experimental or computational means. In the experimental approach, a
protein is selected, and mutated in various positions. Mutants are then
assayed to determine how the various mutations affect function. The effect
on structure is then derived from the effect on function.  A more laborious
approach, but feasible with certain proteins, would be to actually determine
the structure of the mutants. The structural approach is the most direct and
the results given may be considered ``gold standard''.  Experimental
functional studies were the first which suggested that protein function is
maintained by a small number of residues. Studies of the T4 phage lysozyme,
the P22 Arc repressor and the \textit{E. coli} lac repressor are reviewed in
this section. However, experiments can only be performed with proteins for
which functional assays exists. Even so, this is a laborious process, in
terms of work vs. amount of data yielded; and for most proteins, there
exists no functional assay. 

Experimental studies provide a family of related (if artificial) sequences,
complete with bona-fide annotation regarding the role of mutated positions
or subsequences. In a similar manner, evolutionary changes may be studied
using computational methods.  The issues which must be addressed when using
a computational approach are: (1) criteria for the location of critical
residues and (2) verification of results. In cases when the computational
method is used in order to develop a predictive ability, then an assessment
of prediction accuracy should also be provided.

Most computational approaches use evolutionary conservation (although not
exclusively) as a criterion for the detection of key positions. The
rationale is simple: if a position is conserved within a protein's sequence
family, then that is a result of selection, and this position is important
for preservation of the protein's viability.  The methods vary in their
definition of positional evolutionary conservation, and how the results
should be interpreted. For example, Mirny \& Shakhnovich's
conservatism-of-conservatism method differentiates between intra-family and
inter-family evolutionary conservation, drawing conclusions regarding the
function of the different types of conservation \cite{Mirny.Shakhnovich:99}.
Typically, this information is obtained by automatically collecting and
aligning sequence family members from one of the large protein sequence
databases.

Structural information is also incorporated into computational analyses. For
example, positions which are spatially close \textit{and} conserved may
indicate a functional role, if located on the protein's surface, or a
structural role, if buried. All computational studies reviewed here use
structural information, although in different manners. In the 3D cluster
analysis, the predictive function incorporates spatial proximity data
directly in order to assess positional importance. In the
conservatism-of-conservatism and in the CKAAPs study, 3D information is used
for structural alignments during family collection phase, where sequences
are too distant to perform sequence-based alignment.

Following is a review of experimental and computational studies of critical
amino-acid positions. The experimental studies are discussed first.  The
importance of the experimental studies is not only in providing ``gold
standards'' for the location of key positions, but foremostly in showing
that many mutations may be introduced without any deleterious effect on
structure or function.  The computational studies are then provided,
including recent ones.  These are discussed also as a preamble to the
research conducted during the course of my doctorate.

\subsection{Experimental studies}
\label{sec:history_exp}

\subsubsection{T4 Phage Lysozyme}

Rennel \textit{et al.} (1991) have conducted a functional study of 2015
mutations along all 163 positions of the T4 phage lysozyme (T4L) (barring
the initial methionine). They have shown that 55\% of the positions along
the T4L sequence tolerate a minimum of 13 different amino acid substitutions
without any effect of the protein's function. 328 mutations, affecting 45\%
of the residues were scored as deleterious.  Comparing their findings to
T4L's structure, they have found that there is a high correlation between
residue burial and substitution intolerance. Two striking exceptions of
exposed positions were associated with catalytic function.


Critical positions, which were completely intolerant to mutations were found
at the catalytic site, and in two buried salt bridges and their stabilizing
residue network. Two others were found in exposed salt bridges. The total
was 12 positions, \~{}7\% of the protein's length.

Subsequent experiments with the T4L system have been conducted by Brian
Matthews, using functional assays and X-ray crystallography to examine the
sequence-structure relationship in the T4L. In a 1995 review in Advances
in Protein Chemistry, it was concluded that the protein can accommodate
changes in many sites, while still maintaining structure
\cite{Matthews:95}. This robustness is also featured while replacing core
residues, sometimes leading to correct folding through repacking of the
core region. Destabilizing core interactions included leucine to alanine
replacements. The destabilizing effect was attributed to the creation of
cavities leading to destabilization by loss of hydrophobic and
van-der-Waals interactions. In one study up to 10 adjacent residues were
substituted by methionines in the protein's core region
\cite{Gassner_etal:96}. This study has shown that that multiple
replacements with a single amino-acid in the core are possible. Although
eventually leading to a loss of stability, T4L has retained its structural
and functional properties through most of the replacements. Another study,
examining the structure-function relationship has shown that catalytic
site positions may be substituted, leading to a reduction or loss of
activity, but not of structural stability. Stability was actually
increased, suggesting that catalytic site residues are not optimized for
structural stability \cite{Shoichet_etal:95}.

The following tenets were suggested: (1) A subset of the amino
acids in a protein is of key importance for folding and stability. (2)
This subset consists primarily of the interior residues. (3) The role and
importance of a given residue depends on its context within the folded
structure of the protein and can be evaluated by substitution of
alternative amino acids at that site. (4) Catalytic site residues are not
necessarily related to structural stability. In fact, the structure may be
stabilized by mutating the catalytic residues, at the expense of
function loss.

\subsubsection{The bacteriophage P22 Arc Repressor}

Milla \textit{et al.} have studied single alanine-substitutions in the Arc
repressor of bacteriophage P22 \cite{Milla.Brown.ea:94}. The P22 Arc repressor
provides an attractive system for sequence-structure relationships due to
its small size of 52 residues. Fifty-one non-alanine positions were mutated
to alanine in this study.  Using melting temperatures as a measure, it was
shown that twenty-five mutants had $t_m$ values near the wild-type, and 20
mutants were found to be less stable. Five mutations prevented protein
folding altogether, and one mutant (P8A) was found to be more stable than
the wild type.  All mutants were compared with the structure in order to
elucidate the reason for structural disruption, or the lack of it. The
reason for the increased stability of mutation P8A, was given as a relief
of unfavorable packing interactions caused by a buried proline ring. The 25
neutral stability mutations affected side chains which are mostly solvent
exposed, and have high B-factors (see also Markievicz's \textit{et al.}
study below). Eight of the 25 mutations affected residues involved in
hydrogen bonding, while two others affected residues in the hydrophobic
core. The destabilizing mutations mostly affected glycines with positive
$\phi,\psi$ values, hydrogen bonds and salt bridges. Those positions were
found to be more buried and with lower B-factors than positions in the
neutral mutations class. The five mutations which prevented folding
altogether were at positions with low B-factors, four of the mutations
affected hydrophobic core mutations and one a buried polar residue. The
five wild-type side chains altered in the ``unfolded'' mutant class pack
together in the native structure. The authors hypothesized that they may
form a folding core, that may have been disrupted by the mutations.

Brown and Sauer studied mutants of the Arc repressor containing from 3 to 15
partially overlapping multiple-alanine substitutions \cite{Brown_Sauer:99}.
(Five mutants with stretches of 3, 7, 11, 12 and 15 alanines, named 3A, 7A,
etc.) Their choice of positions was purposefully targeted to those positions
which in the previous study were shown individually to have little effect on
protein stability. Twenty-two different residues in total were mutated in
this study. Examinations of the spectral properties of the mutants (using CD
and fluorescence) revealed that they were able to adopt native structures
with many similarities to the wild-type Arc repressor.  Mutants 7A, 12A and
15A were unable to bind DNA, as revealed by footprinting probes. Mutants 3A
and 11A functioned well in that respect. All of the five variants formed
heterodimers with the wild-type Arc. 

In summary, the Arc repressor mutant studies have shown that up to 55\%
of the residues can be mutated individually with no deleterious structural
effect, and 26\% of the positions can be mutated in concert, with the
Arc repressor mutants still assuming the same fold as the wild-type.

\subsubsection{The \textit{E. coli} lac repressor}

Markiewicz \textit{et al.} studied a set of over 4000 single amino-acid
replacements in the \textit{E. coli} lac repressor protein
\cite{Markiewicz.Kleina.ea:94}.  Markiewicz \textit{et al.} have located
segments of tolerant regions, which they replaced with spans of 5 to 13
alanines, preserving repressor function. They concluded that 192 of 328
sites (59\%) are generally tolerant to substitutions. Certain segments were
found to be more restrictive in tolerance, i.e. tolerant of substitutions
within a certain physico-chemical group. A multiple sequence alignment of
the lacI protein with 13 known homologues has revealed a good correlation
between positional conservation and substitution intolerance. Another
phenotype analysis of the lac repressor was conducted
\cite{Suckow.Markiewicz.ea:96} this time with reference to its published
structure. The chief goal was to elucidate structural roles of
substitution-intolerant mutations. In most cases, mutant effects could be
correlated with structural and functional features of the protein. Suckow
\textit{et al.} partitioned the amino-acid positions along the protein into
15 groups, based on their physico-chemical traits, location in the protein,
DNA binding, ligand binding, participation in the dimerization interface,
etc. They have found that the amino acids which are characterized solely by
their solvent exposure, are generally tolerant to substitutions. However,
solvent exposed residues participating in salt bridges were intolerant to
substitutions.  Positions which were composed of small amino-acids were
tolerant to substitutions only by small amino-acids.  Some parts of the
protein, identified as spacer regions, were not only tolerant to single
substitutions, but could completely be replaced with stretched of alanine
residues, placing the alanine stretch replacements observed in the
previous study in a structural context.  Substitutions in positions
participating in DNA contacts, or in ligand binding (isopropyl-$\beta$-{\sc
d}-thiogalactoside, IPTG, which lowers the repressor binding affinity to the
operator DNA by three orders of magnitude), or in the dimerization interface
were not tolerant to substitutions, resulting either in an inactive enzyme,
or in the I$^s$ phenotype (unresponsive to an inducer). Certain
substitutions resulted in the I$^s$ phenotype without being directly
involved in inducer binding. Reasons that were given regarded a failure to
transmit the inducer effect to the DNA binding domain of the protein, or in
some cases, enhancement of DNA binding affinity so that the inducer can not
force the protein to dissociate.

Some positions were identified as directly affecting the protein fold. Those
are positions which are predominantly buried and small (Gly, Ala, Thr and Ser
residues) for which only small amino-acids are tolerated. Small exposed
residues were also found to be intolerant, although the phenotypical effect
was not as drastic as with the small buried ones. 

Proline substitutions affecting secondary structure elements usually occurred
at secondary structure elements involved in dimerization. Thus, even
secondary structure element distortion was occasionally tolerated, or
compensated for.

In conclusion, the lac repressor studies have shown a general robustness of
the protein structure with regards to substitutions. Substitution-intolerant
positions were shown to have a clear structural or functional role.

\subsection{Computational studies}

\subsubsection{Conservatism-of-Conservatism}

Mirny \& Shakhnovich  have studied five of the most populated protein folds
\cite{Mirny.Shakhnovich:99}. They have attempted to separate between
historic, structural and functional reasons for positional conservation.
This was done by comparing intra- and inter-sequence family conservation,
among sequences populating the same fold. In order to do that,
representative structures of selected sequence families which populate a given
fold were chosen, and the following steps were taken:

\begin{enumerate}

\item Construction of multiple sequence alignments of proteins homologous to
each representative protein.

\item Identification of positions which are conserved within each multiple
alignment

\item Structural alignment of families to each other.

\item Identification of sites where conserved positions coincide between the
families. 

\end{enumerate}

The phenomenon of consistent conserved positions amongst families was named
conservatism-of-conservatism, or CoC. The analysis was performed using an
alphabet of six letters, where the amino-acids were grouped by
physico-chemical traits. This was done to factor out interchangeability
which occurs between amino-acids of similar physico-chemical traits.  Two
measures of positional entropy were taken $S(l)$ and $S^{across}(l)$. S(l)
is the measure of CoC between families, and identifies conserved positions
that may vary in residue identity among the families.
$S^{across}(l)$ is the measure of CoC across families, which identified
conserved residue types across all families.

The folds that were analyzed were the immunoglobulin (Ig) fold, the
oligonucleotide-binding (OB) fold, the Rossman fold, and the TIM barrel. The
parameters that were attempted to be correlated with high CoC were
function, thermodynamic stability, and kinetics.
For each of the folds studied, high scoring CoC positions were shown to form
a dense cluster within the native structure. However, cluster location and
participating residues were shown to vary between folds, or within families of
the same fold.  For example, in the Ig fold residues with high CoC form a
cluster deeply buried in the fold. Some families of this fold stabilize this
cluster by hydrophobic interactions, others by disulfide bonds. Generally,
the study revealed a high correlation between CoC and solvent accessibility:
high CoC positions were usually buried. However, in those folds where data
existed regarding folding kinetics, many clusters were shown to form
nucleation centers. In two cases, the analysis of $S^{across}$ revealed the
existence of super-sites, that is a common denominator which can be
attributed to function, regardless of the actual function which the protein
assumes.

\subsubsection{Conserved Key Amino-Acid Positions}

Reddy \textit{et al.} have studied conserved key amino acid positions
(CKAAPs), which were derived from common substructures in proteins from the
entire PDB \cite{Reddy.Li.ea:01}. The subsequences derived from the
substructures are aligned, and the positions in the multiple alignments
score according to conservation.

In this study, several folds were also particularly examined, in order to find
the function of conserved residues according to the CKAAPs method.
Interestingly, one of the folds examined was the Ig fold. Reddy \textit{et al.}
have discovered that the CKAAPs analysis found all the residues found by the CoC
analysis, plus some others.  In contrast with the CoC analysis, the CKAAPs
analysis was performed on many families of the PDB. This gave Reddy \textit{et.
al} the opportunity to select those proteins in which mutations were
well-documented, and compare them with the CKAAPs findings. The well-documented
Arc repressor was examined. It was found that Arc mutations in positions
designated as CKAAPs exhibit more severe perturbations in protein stability. 

In a whole-database analysis, CKAAPs were shown to be no more buried than
the normal pattern of solvent accessibility. However, CKAAPs were found to
be predominant in the terminal regions of rigid secondary structural
elements.  Examination of Ooi numbers shows that CKAAPs are mostly
surrounded by other amino acids and that charged groups on the amino acids
are better neutralized by hydrogen bonding interactions.


\subsubsection{3D cluster analysis}
Landgraf \textit{et al.} used representative structures and multiple
sequence alignments in a method called three dimensional cluster analysis
\cite{Landgraf_etal:01}. In this study, the \textit{regional conservation
score} $C_R(x)$ defines the conservation of each residue and its spatial
neighbors relative to the rest of the protein. A high $C_R(x)$ value means
that position $x$ is located within a conserved spatial cluster within the
sequence family. The \textit{similarity deviation score} $S(x)$ detects
clusters with sequence similarities deviating from the similarities of the
full-length sequences. A high $S(x)$ score indicates a strong deviation
between the similarity relationships within the regional alignment of the
structural neighbors of residue $x$ and the similarity relationships
obtained for the full-length sequences.

The difference between this study and the CKAAPS or CoC studies is that
positions in this study were initially scored based both on their
conservation and their clustering with other conserved positions. The
incorporation of three-dimensional information in that manner enabled
Landgraf \textit{et al.} to detect clusters which cannot be detected simply
by multiple sequence alignment, and which have a potential for controlling
protein function. In this study, 35 different folds were analyzed. The main
goal of the analysis was to evaluate the ability of 3D cluster analysis to
predict functional interfaces, as defined by cocrystal structures. It was
found that $C_R(x)$ identifies the majority of the residues in interfaces.
The identification of interface residues increases as the sequence diversity
within the family increases. The reason being that the signal-to-noise ratio
increases with increased overall sequence diversity: the conserved clusters
remain conserved, whereas the overall sequence identity decreases. 

The $S(x)$ score was used for a different purpose. Here, the question was
whether proteins could possess residue clusters for which the global
sequence similarity relationships might not adequately reflect evolutionary
and functional relationships. Landgraf \textit{et al.} suggest that increased
$S(x)$ scores may indicate regions controlling the specificity of protein
functions.

In addition to the 35 families analysis, 3D cluster analysis was performed
on the MAP-kinase ERK2, and on aldolase. In the ERK2 analysis, $C_R(x)$
identified the P1 site, the ATP-binding pocket, and the dual-phosphorylation
site. There was a considerable overlap between $C_R(x)$ and $S(x)$, however
the highest $S(x)$ scores were exhibited in the specificity-conferring P1
region and the ATP-binding pocket. In the aldolase analysis, the area
showing the highest $C_R(x)$ scores was found to be located in the core of
the $\alpha/\beta$-barrel and includes all key residues known to be involved
in catalysis. 

\subsection{This study's approach}
\label{sec:goals}

This study approaches the problem of critical residue location by studying a whole
set of proteins, without restriction to any given fold or sequence family. The
studied data set is compiled of protein pairs where the pair-mates are widely
different in their sequence, but have the same fold.  The hypothesis being
that there is a cryptic common denominator, present at the sequence level, which
causes such proteins to assume the same fold.  This sequence-encoded common
denominator is not overtly detectable by sequence alignment methods (see section
\ref{sec:seq_hom}), as by definition the collated protein pairs are not
sequence-alignable.

As a first step, a database of structurally similar, sequence dissimilar protein
pairs (SSSD, see \ref{sec:sssd}) was constructed. We then proceeded to locate and
study aligned positions which are suspect of being critical to the fold and
function of the proteins in the database. Initially (Chapter 2) we have looked at
positions which are structurally aligned and identical.  Following that, we studied
positions which are aligned and conserved, using a novel method of detecting
evolutionary conservation in close and distant sequence family members of the
studied proteins (Chapter 3). Computational studies published previous to and
concurrently with this study which aimed to locate critical positions, have used
either a single structure and its sequence family members (3D clustering), or
structural alignment of proteins from the same structure and sequence family (CoC,
CKAAPs). In contrast to those studies earmarked by being family specific, this
study makes a point of examining pairs of proteins which have no detectable
sequence similarity, sifting in only those sequence determinants which
would play a role in assuming the same fold/function for the studied protein
pair.

In another part of this study, we have examined the accuracy of alignments
produced by PSI-BLAST, when compared with the structural alignments we
already had from our database. PSI-BLAST is commonly used to detect
remote homologues. Having a database of structurally similar, sequence
dissimilar protein pairs enabled us to assess the ability of PSI-BLAST to detect
remote homologues, and to evaluate its alignment accuracy.

\section{The Importance of Determining Protein Structure}
\label{sec:structure}

A protein's structure is much more biologically informative than 
sequence only. 
The reasons for that are:
\begin{enumerate}

\item\textsc{Functional:} Structure solution provides knowledge regarding
the biochemical mechanism by which the protein carries out its function. The
roles and mechanisms of elements of a protein become apparent only when
viewed in a structural context. Understanding the mechanism of catalytic
sites, binding sites, protein-protein interfaces, kinetics and
thermodynamics of folding all require a structural solution of the
protein.

\item\textsc{Evolutionary:} Structure is well conserved over evolutionary
time, and thus it enables the recognition of evolutionary relatedness with
other proteins, undetectable by sequence comparison.

\end{enumerate} 

Automated sequencing techniques have inundated us with protein sequences.
Even before the coordinated genomic sequencing efforts, large repositories
of sequence data have been created simply by the contributions stemming from
ongoing research. There are currently \~{}600,000 protein sequences in the
protein sequence databases SwissProt and TREMBL \cite{Bairoch_Apweiler:00}.
However, with regard to protein structure, the amount of data we have is
much smaller. The reasons for that are: (1) the lower throughput of
structure determination methods: X-ray crystallography and NMR; (2)Structure
determination methods are currently limited almost exclusively to globular
proteins; (3)Even among those proteins there are certain proteins which
cannot be crystallized, or are too large for NMR spectroscopy.

As a result of those limitations, targets for structure determination were
chosen much more selectively than targets for sequence determination. When
compared with sequencing, structure determination is a costly and lengthy
undertaking. The Protein Data Bank
(\textbf{PDB})\cite{Berman.Westbrook.ea:00} currently holds
\~{}16,000 structures.  

The importance of producing structural solutions is exemplified by the
recently launched effort of \textbf{structural
genomics}\cite{Brenner:01}. Structural
genomics aims to provide tractable solutions to all proteins, by creating
a library of solved representative structures, and using computational
means to assign a fold to all known sequences.

\section{Methods and Materials}
\label{sec:methods}
This section explains, and when necessary elaborates upon, the methods used
in this study.

\subsection{Homology by sequence, and sequence families}
\label{sec:seq_hom}

The most common way of inferring and studying an evolutionary relationship
between proteins is by sequence alignment. There are currently several
algorithms which enable us to align two or more protein sequences, in order
to determine their similarity. Pairwise alignment, the alignment of two
sequences, is commonly performed using the Smith-Waterman
\cite{Smith.Waterman:81} or Needleman-Wunsch \cite{Needleman.Wunsch:70}
algorithms. Pairwise alignment of protein sequences is initially performed
in order to determine whether they are \textbf{homologous}, that is,
originating from a common ancestor. Homology which is a qualitative trait, is
inferred from the degree of similarity, a quantitative trait, between the
sequences. Inferral is performed by setting a threshold on the degree of
similarity between the sequences, beyond which homology is considered to be
established. Once homology has been established, additional questions may be
asked and answered based on the examination of the alignment. For example,
are the sequences homologous throughout their length, or in some conserved
local region? Are functional areas preserved? 

Dynamic programming is used for pairwise sequence alignment. The rationale
is that the optimal solution for the alignment of the two sequences stems
from the optimal solution of the previous alignment step. Using recursion,
it is possible to establish the optimal shortest distance between any two
sequences.  The score of the alignment achieved can be assessed by placing
it within a distribution of mean random alignments, establishing a
statistical significance for the alignment score.

\subsubsection{Amino acid substitution matrices}
\label{sec:subs_mat}

Amino acid substitution matrices such as the PAM \cite{Dayhoff:78} or
BLOSUM \cite{Henikoff.Henikoff:92} series are typically used in the process
of protein sequence alignment.  When there is call for an amino-acid
replacement (as opposed to an insertion/deletion event), the cost of this
replacement is assessed using the substitution matrix. Each entry in a
\textbf{log-odds} substitution matrix contains the following value:

\begin{equation}
M_{i,j} =  \lambda/\log_2(\frac{P_{i,j}}{P_iP_j})
\end{equation}

Where $i$ and $j$ are any two given amino acid types, or the same one.
$P_{ij}$ is the observed frequency of replacements between $i$ and $j$ in
the database, $P_i$ and $P_j$ are the respective probability of occurrence
for amino acids $i$ and $j$ in the database. Thus ${P_iP_j}$ is the
probability of a random replacement between $i$ and $j$. This is placed on a
logarithmic scale, typically base 2, and the resulting score $M_{ij}$ is
said to be expressed in \textbf{bits}.  $\lambda$ is a scaling factor. 

Consider the BLOSUM series of matrices: each BLOSUM{\em X} matrix is
composed of a subset of aligned sequences with $\leq${\em X}\% identity from
the BLOCKS database of multiply aligned sequences. Thus BLOSUM85 contains
the log-odds replacement frequencies for all aligned sequences in BLOCKS
with a 85\% identity and less. These replacements may be compared to the
ones exhibited in BLOSUM35 (constructed from BLOCKS of 35\% identity or
less), for example, and conclusions may be drawn regarding the differential
frequency of replacements given different evolutionary distances. 

Substitution matrices may also be used for the analysis of databases of
aligned proteins. An important measure which may be drawn from a
substitution matrix regarding the set of alignments it represents is its
\textbf{relative entropy}. A matrix's $M$ relative entropy is calculated as:

\begin{equation}
H(M) = \sum_{i\leq j}^{20}P_{ij}\log_2({\frac{P_{ij}}{P_iP_j}})
\end{equation}

Relative entropy provides a measure of how stringent or promiscuous are the
substitutions in the alignments represented by the matrix. A larger relative
entropy represents a more stringent distribution of substitutions. BLOSUM85
has a relative entropy of 1.085 bits, whereas BLOSUM35 has a relative
entropy of 0.34 bits. The reason is that the smaller the evolutionary
distance between the proteins, the less frequent are the non-synonymous
replacements. 
Thus, studying a matrix's relative entropy, and comparing it with those of
other matrices, is a good measure by which to evaluate substitution
strictness within a data-set. Of course, the matrix should also be studied
in a particular manner, in order to get detailed results.

\subsubsection{Multiple sequence alignment, and sequence families}

Several protein sequences may be aligned together, to form a
\textbf{multiple sequence alignment}. 
The considerations for multiple sequence alignment are 
the same as that for pairwise sequence alignment, except for the major
caveat that dynamic programming cannot be used due to its time and memory
inefficiency. Therefore, heuristics such as the Clustal
\cite{Thompson_etal:94} algorithm are used, based on a progressive alignment
of pairwise alignments.

Using combinations of pairwise or multiple sequence alignment techniques, it
is possible to map the known protein sequence space into families. Indeed,
several databases containing sequence families, of whole proteins or of
protein domains exist. In the recent Nucleic Acids Research database
issue (January, 2002) 18 such databases were listed.
(http://www3.oup.co.uk/nar/database/cat/12). Although some of the databases
listed there deal solely with sequence motifs, most of them are a
compilation of sequence families. A \textbf{sequence
family} is loosely defined as a collection of sequences
which can be aligned in a significant manner, and which have a functional
and evolutionary relationship. The definition is considered loose, as
varying alignment definitions and thresholdings can serve to create
different families from the same pool of sequences.

\subsection{Homology by structure, and structure families}
\label{sec:struct_hom}

Another way of inferring a relationship among proteins is by structural
considerations. This is considered to be a much stronger approach than
sequence-based approaches, as structure is better conserved than sequence
\cite{Lesk_Chothia:80}.
It is used less often, however, due to the paucity of structures available,
and until recently, the paucity of structural alignment programs, and the
lack of accessibility to those which do exist.  There are several automated
structural alignment algorithms e.g.,
\cite{Orengo.Taylor:96,Holm_Sander:93,Leibowitz_etal:01}, for comprehensive
reviews and assessments see \cite{Godzik:96,Gerstein_Levitt:98}. Automated
structural alignment tools are an important part of bioinformatics research,
and more so once it has been realized that many proteins which share a
common fold have no detectable sequence similarity. As stated in section
\ref{sec:structure}, most of the information we have regarding proteins is
sequence information, whereas only for a minority do we have solved
structures. However, we do have enough structure solutions so that we can
examine relationships among proteins based on structural considerations.
That is, superimpose whole proteins, or \textbf{structural
domains} using \textbf{structural
alignment} techniques, in order to discover the
relationship among them. Structural domains are defined as autonomous
folding units within a single protein chain, which are usually associated
with a given function.

The representation of a structure ---3D information--- is much more complex
than sequence 1D information. Consequently, The alignment of two structures
is quite a complex problem. The problem is compounded by the lack of a
representative model for structure, and consequently of a single model of
similarity for structure alignment. Regardless of the method employed, all
sequence alignment methods assume that the sequences to be aligned have a
common ancestor, and that the two sequences are related to each other by a
series of quantifiable steps, which are \textbf{indels} (insertion and
deletions of sequence elements) and substitutions. The methods vary in
algorithmic application (various heuristics vs. dynamic programming), and
indel/substitution penalties applied. Furthermore, the representation of
sequences is always performed using a string of characters. In contrast,
there is no requirement for an underlying evolutionary hypothesis for
structural alignments. Indeed, some alignment algorithms, such as the
geometric hashing,  rely on rigid body superimpositions, and are sequence
independent \cite{Nussinov_Wolfson:91}.

Different algorithms can be used for structural alignment. Among those used
are graph theory, geometric hashing \cite{Bachar_etal:93}, distance plot
comparison \cite{Holm.Sander:96}, and double-dynamic programming
\cite{Orengo.Taylor:96}. Graph theoretical approaches are based on the
representation of the protein as a graph, e.g. using the distance and angles
between secondary structure elements, the latter being the nodes of the
graph. Partial matches between proteins are performed using sub-graph
isomorphism algorithms. Geometric hashing, introduced as a method for
structural alignment by Nussinov \& Wolfson (1991) 
decomposes the proteins to be aligned, representing them as partial rigid
objects. The combinatorial extension (CE) algorithm uses aligned fragment
pairs (AFPs) \cite{Shindyalov_Bourne:98}. AFPs are pairs of fragments from
the proteins to be aligned. Combinations of AFPs which represent possible
continuous paths are selectively extended or discarded, eventually producing
a single optimal alignment.

In the course of my work I have used two different structural alignment
methods. Those were the DALI \cite{Holm.Sander:96} and SSAP
\cite{Orengo.Taylor:96} algorithms.  Those methods were chosen as they have
been producing credible results, for their flexibility and ease of use.
Results produced in this work as a result of
both alignments do not differ significantly from each other (for details on
this see Chapter 2). Both structural alignment algorithms shall be briefly
described here.

The DALI (Distance-matrix ALIgnment) algorithm uses a \textbf{distance
matrix} representation of a protein structure, which is obtained as follows:
a matrix is drawn, with the protein sequence running along the top and side
of the matrix. For each pair of residues that are determined to be in
contact (below a certain distance threshold), a mark is made in the cell
representing those two residues. It follows that for two structurally
similar proteins, two similar distance matrices will be drawn. In order to
align them, the distance matrices are first decomposed into elementary
contact patterns, e.g.  hexapeptide-hexapeptide submatrices. Then, similar
contact patterns in the two matrices are paired and combined into larger
consistent sets of pairs.  This methods allows for the accumulation of
indels.

SSAP (Sequential Structure Alignment Program) uses the Needleman-Wunsch
algorithm originally developed for sequence alignment. However in this case,
three-dimensional geometry is compared to identify equivalent positions.
This is done as follows: 

\begin{enumerate}

\item Construct a view for each residue. The view is a set of vectors from
each C$\beta$ atom to C$\beta$ atoms of all other residues in the protein. A
common frame of reference is defined for each residue based on the
tetrahedral geometry of the C$\alpha$ atom. 

\item The optimal pathway aligning the views is obtained, similarly to
sequence alignment, by dynamic programming. \end{enumerate}

Alignment of views is suggested to be more informative than the alignment of
distance plots (e.g. DALI), vectors give more information on relative
positions than distances. Similarly, C$\beta$ atoms give more information
than C$\alpha$ atoms.

As in protein sequence space, families of protein structures based on their
structural similarities may be obtained. It follows that a map of protein
structure space can be generated using structural alignment techniques, or a
combination of sequence and structure alignment techniques. The three best
known and often used maps are SCOP \cite{Murzin.Brenner.ea:95}, CATH
\cite{Orengo.Michie.ea:97} and FSSP \cite{Holm.Sander:96}. SCOP and CATH
contain a hierarchical representation of the structural information in the
PDB. FSSP is constructed differently, and contains a clustered
representation of protein structures. 

\subsubsection{Structural databases} 

Although seemingly few when compared with the number of sequences, there is
a need to order and classify the \~{}16,000 structures resident in the PDB.
Structural classification is the initial step necessary for recognition of
evolutionary and functional relationships among proteins.

The three main structural classification databases are reviewed here. The
hierarchical databases, CATH and SCOP partition the protein structure space
in a hierarchical fashion. CATH clusters proteins at four major levels,
Class(C), Architecture(A), Topology(T) and Homologous superfamily (H)
\cite{Orengo.Michie.ea:97}. Class, derived from secondary structure content,
is assigned for more than 90\% of protein structures automatically. The
partitioning is into all-alpha helix structures, all-beta strand structures,
and alpha+beta. Architecture, describes the gross orientation of secondary
structures, independent of connectivities. The topology level clusters
structures according to their topological connections and numbers of
secondary structures. The homologous superfamilies cluster proteins with
highly similar structures and functions.  The assignments of structures to
topology families and homologous superfamilies are made by sequence and
structure comparisons.

SCOP (Structural Classification of Proteins) is partitioned in a similar
manner as CATH. However, the structure curation and classification is
performed manually \cite{Murzin.Brenner.ea:95}.

FSSP (from which SSSD, the database in this work is derived, see section
\ref{sec:sssd}) is described here. FSSP stands for \textit{F}old
classification based on \textit{S}tructure-\textit{S}tructure alignment of
\textit{P}roteins \cite{Holm.Sander:96}. FSSP is constructed as follows: 

\begin{enumerate}

\item Representative sequences from the PDB are chosen. The sequences are
chosen so that there is no more than a 25\% similarity between any two
sequences. 

\item The representative sequences are structurally aligned, using DALI, to
all other PDB structures.

\item For each representative structure, an entry in the FSSP database is
created. Each entry contains the alignment of that structure to all PDB
structures for which a statistically significant alignment exists, according
to DALI.

\end{enumerate}

DAPS is a subset of FSSP \cite{daps:01}, which contains alignments from
those entries which have a low sequence identity percentage (25\% or less).

\subsection{The SSSD database}
\label{sec:sssd}

A central implement used for my work was a rigorously compiled database of
118 structurally similar, sequence dissimilar protein pairs. (The SSSD
database). The database is composed of pairs of proteins whose
structures have been determined by X-ray crystallography. The structural
alignment of these proteins has been determined by the DALI algorithm, and
they have been extracted from the FSSP and DAPS databases. The SSSD database
has the following traits:

\begin{enumerate}

\item A minimal protein length of 30 residues. The reliability of the
structural solution of shorter peptides is often dubious, and they are more
prone to radical structural changes based on single-residue substitution.

\item Each protein has been determined by X-ray crystallography with a
resolution of $\leq 3.5$ \AA. This ensures that a reasonably accurate
structural solution was obtained. 

\item Difference in pair member lengths does not exceed 50\% of the shorter
member's length, and is at least 60\% of the longer pair-member's length.
These criteria ensure a large mutual overlap length of the alignments, so
that the structural alignment would be of at least one structural
domain.

\item The Smith-Waterman algorithm was used to check the statistical
significance of pairwise sequence alignment of the pair-mates. Pairs whose
local alignment was found to be statistically significant, were excluded
from the database. Therefore, the pair-mates in the SSSD database do not
have any detectable sequence similarity.

\end{enumerate}

Taken together, these traits ensure that SSSD contains pairs of proteins
which are well-aligned structurally, but have no sequence similarity between
the pair-mates. The pairwise structural alignments in SSSD were used as the
starting point for the location and characterization of positional
determinants which are suspected of being important for protein fold /
function. Initial analysis of SSSD shows that the distribution
of folds within it parallels that of the entire PDB. 

\subsection{PSI-BLAST}
\label{sec:psi_blast} 

Position Specific Iterated Basic Local Alignment Search Tool
\cite{Altschul.Madden.ea:97} (\textbf{PSI-BLAST}) is a
program used for searching sequence similarities in protein databases.
Briefly, PSI-BLAST works in a series of repeated iterations. First, a
protein sequence (the \textit{query sequence}) is given to the program.
PSI-BLAST searches for sequence similarities in whatever database the user
specifies. Second, the results of the search are aligned, and a position
specific scoring matrix (\textbf{PSSM}) or profile is
generated. The PSSM is a matrix which indicates the frequency of each amino
acid in each aligned position.  Thus, for each position in the alignment,
the PSSM contains information both about the amino-acid types in that
position, and the overall conservation of the position. In the third step,
the database is searched again with the PSSM. Searching with a PSSM instead
of searching with a query sequence sensitizes the search to include more
sequences, as the PSSM contains information from the alignment of several
sequences, thus representing a sequence family rather than a single
sequence. The second and third steps are reiterated, for a predetermined
number of iterations, or until the search converges: no new sequences are
found in the queried database. In my work I have used PSI-BLAST as a tool
for determining positional conservation along protein sequences (chapters 2
\& 3). The positional frequency of amino-acids in PSI-BLAST is not readily
available, and I was required to modify PSI-BLAST's source code in order to
extract that information.
One important parameter in PSI-BLAST, which is referred to in Chapters
\ref{chap:mpc}, \ref{chap:discussion} is the e-value.
The \textbf{e-value} (``Expect-value'') is a parameter that describes
the number of hits one can ``expect'' to see just by chance when searching a
database of a particular size. Essentially, the e-value describes the random
background noise that exists for matches between sequences.  The e-value is used as
a convenient way to create a significance threshold for reporting results.
\subsection{Materials}

Most of the code independently written for this study was developed in Python
(python.org), using the the Biopython toolkit (biopython.org). A minority was
developed in C.  Computations were performed on RH Linux 6.2 and 7.1 (Red Hat
Inc.) Intel 686, and on Silicon Graphics (Silicon Graphics, Inc.) Indy IRIX 6.5.
Databases used (CATH, FSSP, DAPS, NCBI non-redundant) for this study were
downloaded from their respective sites, with updates and citations noted in the
papers presented. The standalone version of PSI-BLAST, blastpgp, was used for
PSI-BLAST runs. I have modified the source code of blastpgp in order to obtain
observed and expected residue frequencies making up the PSI-BLAST PSSMs.

Database parsers and program output parsers were developed independently, or
taken from the Biopython toolkit. Program suites such as GCG 10.1 (Accelrys,
Inc.) and EMBOSS \cite{Rice_ea:00} were occasionally used for the processing
of sequence data. 

\section{The Challenge of Fold Prediction}
\label{sec:chal_fold_pred}

Fold prediction is the technique of predicting a fold for a given sequence which
shares no similarity with other sequences in the database. Methods of fold
prediction vary. {\bf Threading} uses pseudo-energy functions to determine
whether a sequence can be aligned using energy considerations, with any of the
known folds. PSI-BLAST, and more sophisticated methods derived from PSI-BLAST,
may be described as extremely sensitive sequence detectors, and use PSSMs to
detect remote sequence similarities, which may be used for modeling. In any
case, these fold-prediction techniques can rarely predict new folds. Ab-initio
methods try to predict the fold given energy considerations alone, so
hypothetically they may be able to predict novel folds. However, they
are currently quite impractical, both in terms of computational time, and in
success in target prediction.

Much effort in fold prediction is given to the training and refinement of
the techniques. This work is not directly concerned with prediction, but
rather with understanding the mechanisms leading to a given fold.
Specifically, with the detection and characterization of specific positions
which can serve as anchors for folding and function. The ability to locate
and understand the role of these positions can improve the construction of
fold predictors. 

\section{Synopses of following chapters}

Chapter 2 is a published paper describing the study of identical, aligned
conserved residues between the SSSD pair-mates. It shows that among
structurally aligned protein pairs identical residues which are conserved in
evolution tend to be located in buried positions, and many are found in
positions critical for maintenance of structure or function.

Chapter 3 is a published paper describing the study of all aligned,
conserved residues between SSSD pair-mates. This paper introduces the novel
concept of \textbf{persistent conservation}, that is, an assessment of
conservation based on close and distant sequence family members. A
significant fraction of these mutually, persistently conserved positions
(\textbf{MPC}s) are shown to be located in positions which
are conducive to secondary structure determination, are mostly buried, and
many of them form spatial clusters within their protein structures. A
substitution matrix based on MPCs shows distinct characteristics which may
prove valuable in protein design experiments.

Chapter 4 is a published paper which examines the accuracy of PSI-BLAST
alignments \textit{vs.} structural alignments. Notably it shows improvement
in alignment sensitivity over consecutive iterations with no discernible
specificity loss. It also shows that in terms of alignment accuracy,
PSI-BLAST performs as well as threading methods.

\chapter{Paper:
Glimmers in the Midnight Zone: Characterization of Aligned
Identical Residues in Sequence-Dissimilar Proteins Sharing a Common Fold}
\label{chap:stairs}

\small
Friedberg I, Kaplan T and Margalit H
\textit{Proc Int Conf Intell Syst Mol Biol.} (2000) 8:162-70.
\normalsize

\chapter{Paper: 
Persistently Conserved Positions in Structurally-Similar, Sequence
Dissimilar Proteins: Roles in Preserving Protein Fold and Function}

\addtocounter{page}{9}
\label{chap:mpc}

\small 
Friedberg I and Margalit H
\textit{Protein Science} (2002) 11(2):350-60
\normalsize


\chapter{Paper: 
Evaluation of PSI-BLAST alignment accuracy in comparison to structural
alignments}
\label{chap:psi-blast}
\addtocounter{page}{11}

\small 
Friedberg I, Kaplan T and Margalit H
\textit{Protein Science} 2000 9(11):2278-84.
\normalsize

\input{discussion_arxiv.tex}

\begin{singlespace}
	\bibliographystyle{bioinformatics}
	\bibliography{iddo_ref_arxiv}
\end{singlespace}

\end{document}

%% file: ack.tex
I owe the completion of this work to quite a few individuals:

To my parents, Devorah and Ilan who taught me the love of science.

To my wife, Vered, who has lovingly supported me during the inevitable hardships
that a doctoral program entails.

To Hanah Margalit, my advisor. Hanah's unique blend of patience, optimism, high
working standards, originality, and scientific critique have made this work
possible.  I believe that if I am a better researcher now than I was five years
ago, it is solely due to her guidance.

To Tommy Kaplan, who helped in producing and collating information from the SSSD
database. An extraordinarily gifted person, and I am glad to have had the
opportunity to work with him.

To the people involved in the Biopython project, a set of open-source tools for
bioinformatics to which I have contributed little, and from which I have taken
much. This fine package has saved me months of coding time.  I would especially
like to thank Brad Chapman and Jeff Chang for introducing me to the finer points of
open-source collaboration. 

To the people in Hanah Margalit's group, who have provided a warm work atmosphere,
and a helpful professional environment.

And finally, to my children, Nitzan and Neta, who have brought unimaginable joy
to my life.

%% file: summary_arxiv.tex

Mapping between sequence and structure is currently an open problem in
structural biology. Despite many experimental and computational efforts it is
not clear yet how the structure is encoded in the sequence.  Answering this
question may pave the way for predicting a protein fold given its sequence.

My doctoral studies have focused on a particular phenomenon relevant to the
protein sequence-structure relationship. It has been observed that many
proteins having apparently dissimilar sequences share the same native fold.
The phenomenon of mapping many divergent sequences into a single fold raises
the question of which positions along the sequence are important for the
conservation of fold and function in dissimilar sequences. Detecting those
positions, and classifying them according to role can help understand which
elements in a sequence are important for maintenance of structure and/or
function.  In the course of my doctoral research I have attempted to
discover and characterize those positions. 

The method I undertook was as follows: I constructed a database of
structurally similar, sequence dissimilar protein pairs, as a tool for
detecting those positions. Aligned positions between pair-mates were
examined for evolutionary conservation within their respective sequence
families. Positions having a mutual conservation in the non-intersecting
sequence families of each pair-mate are deemed to play a role in the
conservation of fold and/or function between aligned pair-mates. The
rationale being, that evolutionary conservation between aligned positions is
due to the preservation of some critical aspect for fold or function.

In the initial phase of this work (Chapter \ref{chap:stairs}), I have examined those
positions which were structurally aligned, and possess identical residues.
It was shown that out of Structurally Aligned, Identical ResidueS (STAIRS),
40\% are only moderately conserved, suggesting that their maintenance as
identical residues was coincidental. However, STAIRS with high mutual
evolutionary conservation exhibit low solvent accessibility, and an
over-representation of certain amino-acids. We also examined a subset of
STAIRS which are spatially proximal (neighboring STAIRS or NSTAIRS).  An
itemized examination of the over-represented STAIRS which are spatially
proximal has shown that an overwhelming majority can be assigned with a
functional or a structural role: location in functional sites, and
determination of secondary or super-secondary structure.

The evolutionary conservation of positions was determined automatically by
multiple alignment of each sequence to homologues in the NCBI non-redundant
sequence database using the PSI-BLAST program. PSI-BLAST was used to
determine the conservation in each position over several iterations, in a
way which enabled the expansion of the ``evolutionary horizon'', as each
iteration provides alignments with sequences which are more evolutionary
distant. I then proceeded to look at those positions which are {\bf
persistently conserved} in each pair-mate, and in both aligned structures.
Persistently conserved positions have been defined as those positions which
are conserved both at the first and last PSI-BLAST iterations for the
following reason: conservation in only the first iteration might be due to
evolutionary non-divergence, whereas conservation only in the last iteration
might be due to a drift which sometimes occurs in PSI-BLAST, where the
alignment generated by PSI-BLAST no longer holds information pertaining to
the original query sequence. The intersection of the first and last
iterations however, yields positions which are suspect of being conserved
for a structural or functional reason. We have shown that few discreet
positions are conserved in each pair.  Those positions, dubbed MPCs
(``Mutually, Persistently Conserved positions''), were shown to play a role in
helix stabilization, hydrophobic core formation, and active sites.  MPCs
tend, in many cases, to form spatial clusters within their protein
structures.  We have derived a substitution matrix from the MPCs, which we
believe might be conducive to protein engineering and design applications
(Chapter \ref{chap:mpc}).

The database of SSSD protein pairs has allowed us to assess the accuracy of
alignments produced by PSI-BLAST.  Alignment accuracy is important for the
correct modeling of a sequence by the structure of a homologue. PSI-BLAST is
often used for the detection of query sequence homologues, with a solved
structure. In the SSSD database, several pair-mates are detected uni- or
bi-directionally using PSI-BLAST, in the second or following iterations.
Having that data enabled the comparison of alignments produced by PSI-BLAST
to the structural alignments, the latter being the ``gold standard'' for
alignment evaluation. I have shown that from the 123 structurally similar,
sequence dissimilar protein pairs, 52 pairs have detected their pair-mates,
and for 16 of those, the detection was bi-directional. The alignment
specificity was shown to be \~{}44\%, and it does not improve significantly
over consecutive iterations. The alignment sensitivity was shown to be
\~{}51\% at best, and shows improvement over several iterations (Chapter
\ref{chap:psi-blast}). Based on these findings we concluded that the alignment accuracy
produced by PSI-BLAST is as good as those produced by various threading
algorithms, and that the alignment sensitivity may be improved over
consecutive iterations.


%% file: discussion_arxiv.tex

\chapter{Additional Results}
\label{chap:add_res}
\addtocounter{page}{7}

This chapter elaborates upon another case study, in the same manner as in Chapter
\ref{chap:mpc}. The protein pair chosen from the database was Methionine synthase
(MetS, PDB:1BMTA) \cite{Drennan_etal:94} and CheY (PDB:3CHY)
\cite{Volz_Matsumura:91} from \textit{E. coli}.

\section{Description of the Enzymes}

\subsection{Methionine Synthase}

MetS is composed of two domains, a Rossman fold,
the C-terminal domain, and an orthogonal $\alpha$-helix bundle, the N-terminal
domain. Sandwiched between the two domains is the cobalamin prosthetic group. Free
methylcobalamin, a derivative of vitamin B12, is composed of a heme-like corrin
ring, and a dimethylbenzimidazole nucleotide moiety. In the center of the corrin
ring lies a cobalt atom, that is hexacoordinated by four nitrogen ligands provided
by the corrin macrocycle, a methyl group in the upper ($\beta$) axial position, and
a nitrogen (N3) from dimethylbenzimidazole in the lower ($\alpha$) axial position.
However, when bound to Methionine synthase, the dimethylbenzimidazole nucleotide is
displaced from the cobalt to form a ``nucleotide tail''. The corrin portion is
sandwiched between the two domains of the protein, while the nucleotide tail
penetrates into a pocket in the Rossman fold domain (\ref{fig:mets_chey} a).

MetS catalyzes the synthesis of methionine by two successive methyl transfers:

\begin{enumerate}

\item CH$_3$-cob(III)alamin + homocysteine $\rightarrow$ cob(I)alamin + methionine

\item cob(I)alamin + methyltetrahydrofolate $\rightarrow$ CH$_3$-cob(III)alamin +
tetrahydrofolate

\end{enumerate}

\subsection{CheY} 
\begin{figure}[h]
\begin{center}
	{\bf a}
	\includegraphics[width=7cm]{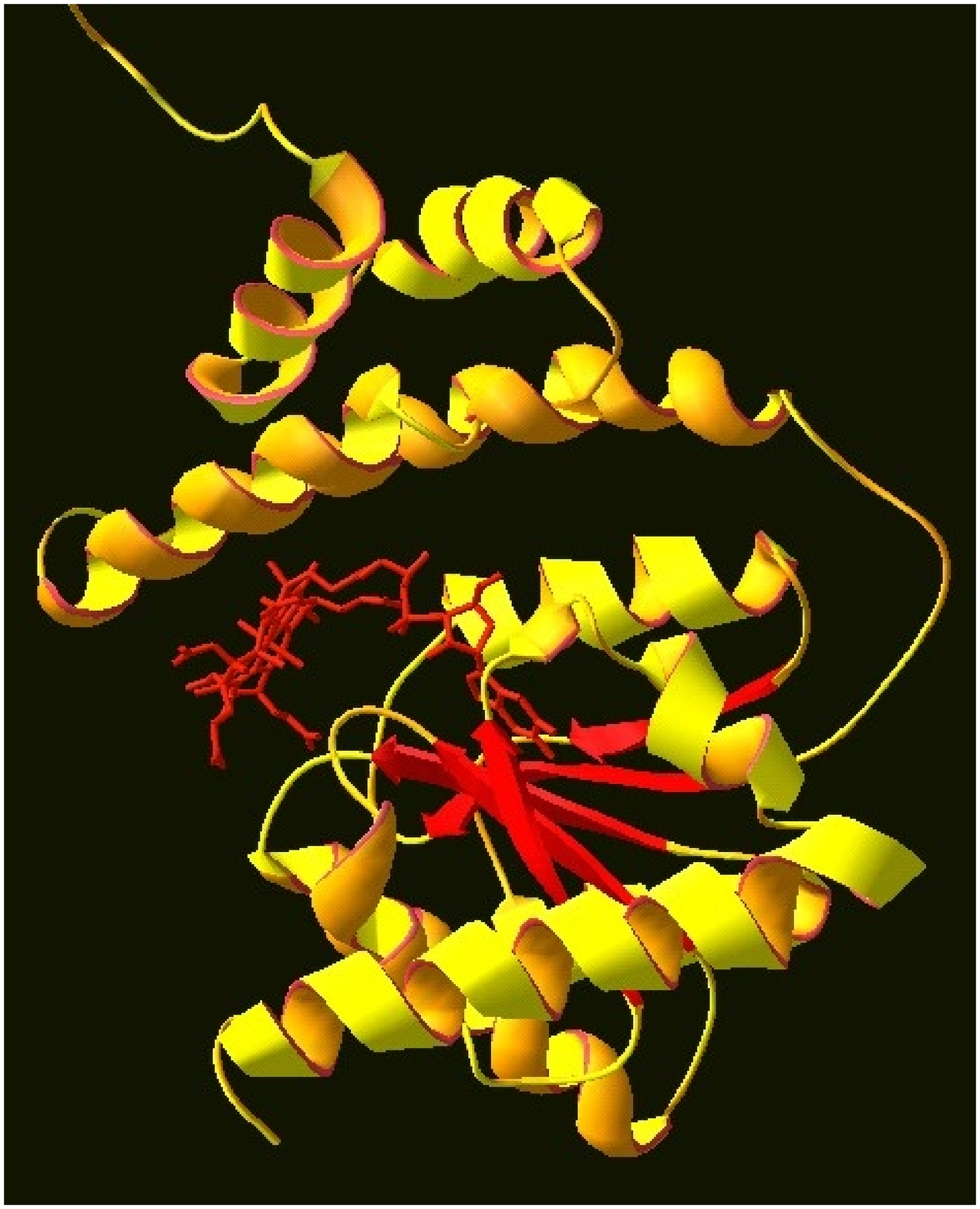}
	{\bf b}
	\includegraphics[width=5cm]{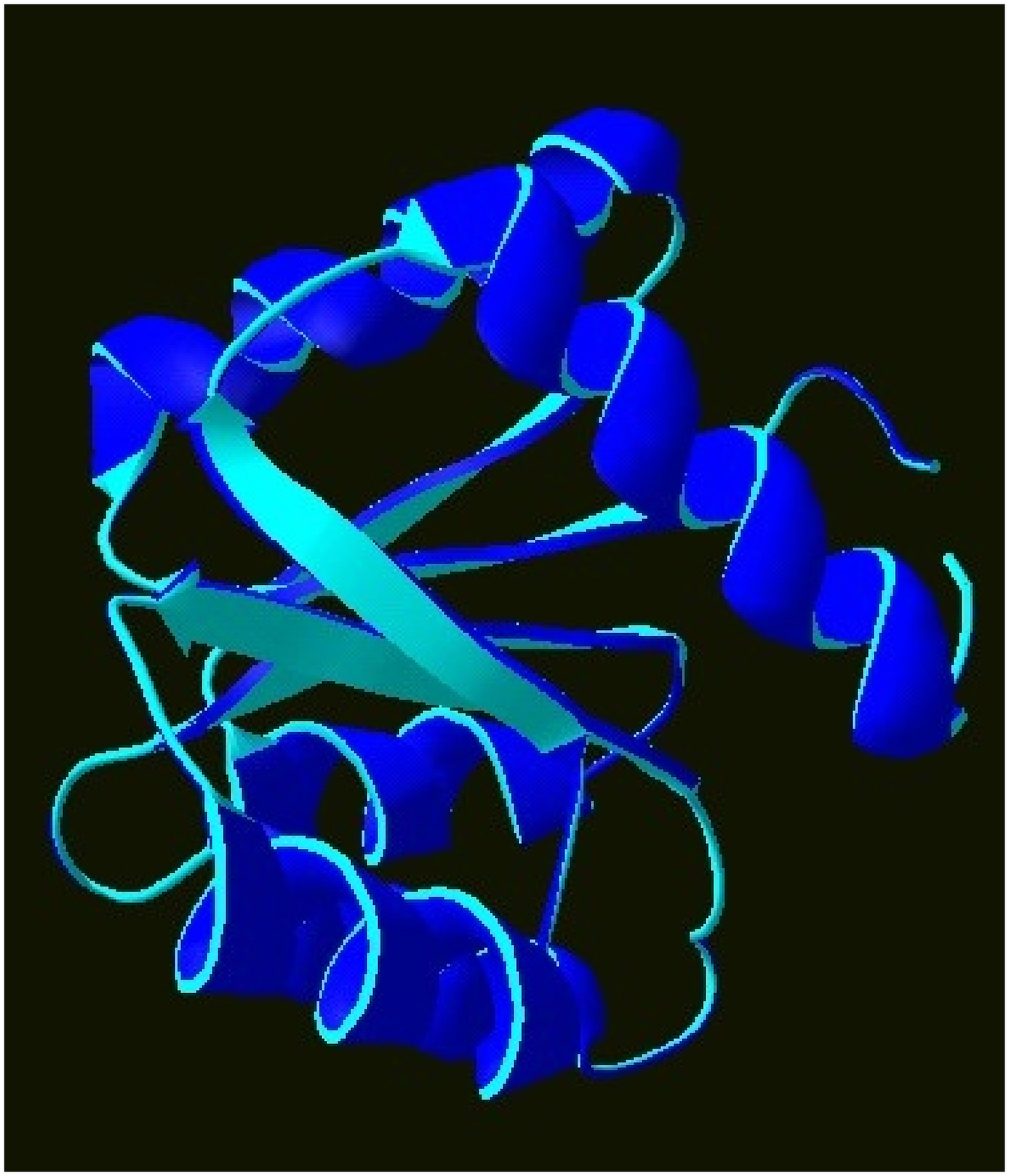}
	\caption{\small
	         (a) yellow: MetS; red, stick representation: cobalamin prosthetic group;
				red, ribbon representation: $\beta$-sheet of the Rossman fold.
	         (b) blue: CheY; cyan: $\beta$-sheet of the Rossman fold}
	\label{fig:mets_chey}
\end{center}
\end{figure}

CheY is a single domain protein, assuming a Rossman fold (Figure
\ref{fig:mets_chey} b).  CheY is part of a chemotactic signal transduction mechanism
in the \textit{E.  coli} bacterium. CheA$_L$, a kinase of the chemotaxis system,
phosphorylates CheY which acts as the response regulator. Transiently phosphorylated
CheY interacts with the bacterial flagellar motor to cause clockwise rotation of the
flagella, creating a distinct motion response dubbed ``tumbling''.

\section{The Alignment}

\begin{figure}[h]
\begin{center}
	\includegraphics[width=10cm]{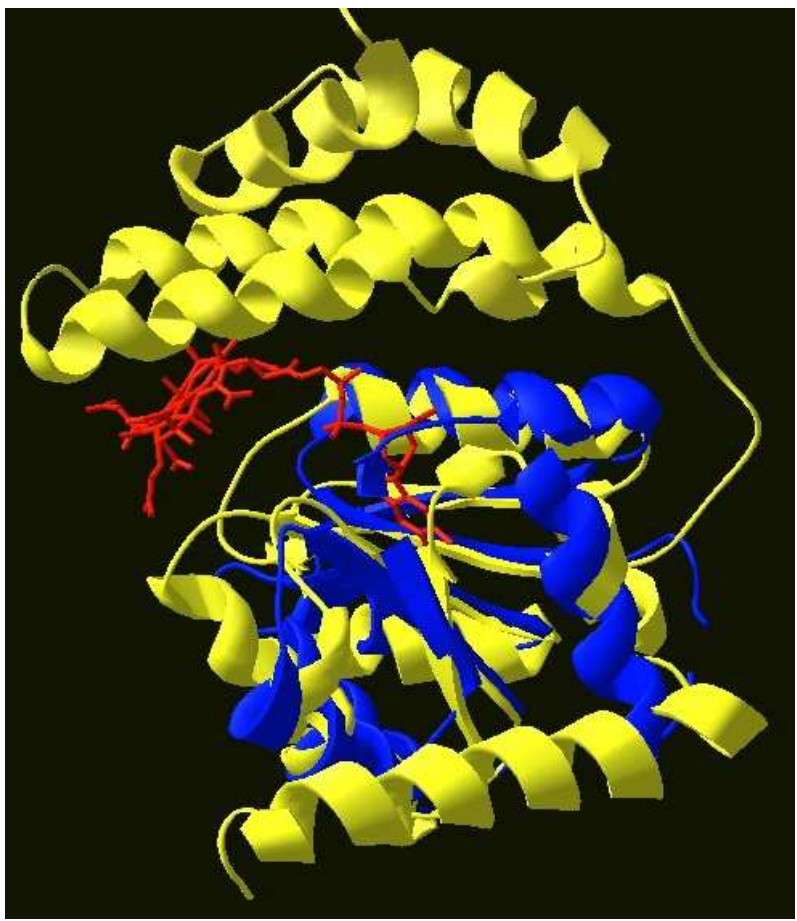}
	\caption{\small Structural alignment of MetS and CheY, as performed by DALI.
	Yellow: MetS; red: cobalamin prosthetic group; blue: CheY}
	\label{fig:mets_chey_sup}
\end{center}
\end{figure}

MetS and CheY are two enzymes that on a sequence and functional level have nothing
in common. The former is a methyltransferase, while the latter is a phosphorylated
signal transducer. However, the Rossman-fold domain of MetS bears a strong
structural similarity to that of CheY. When aligned, (figure
\ref{fig:mets_chey_sup}) several MPCs which have distinct but different functions in
both proteins are revealed. 

\begin{figure}[h]
\begin{center}
	{\bf a}
	\includegraphics[width=8cm]{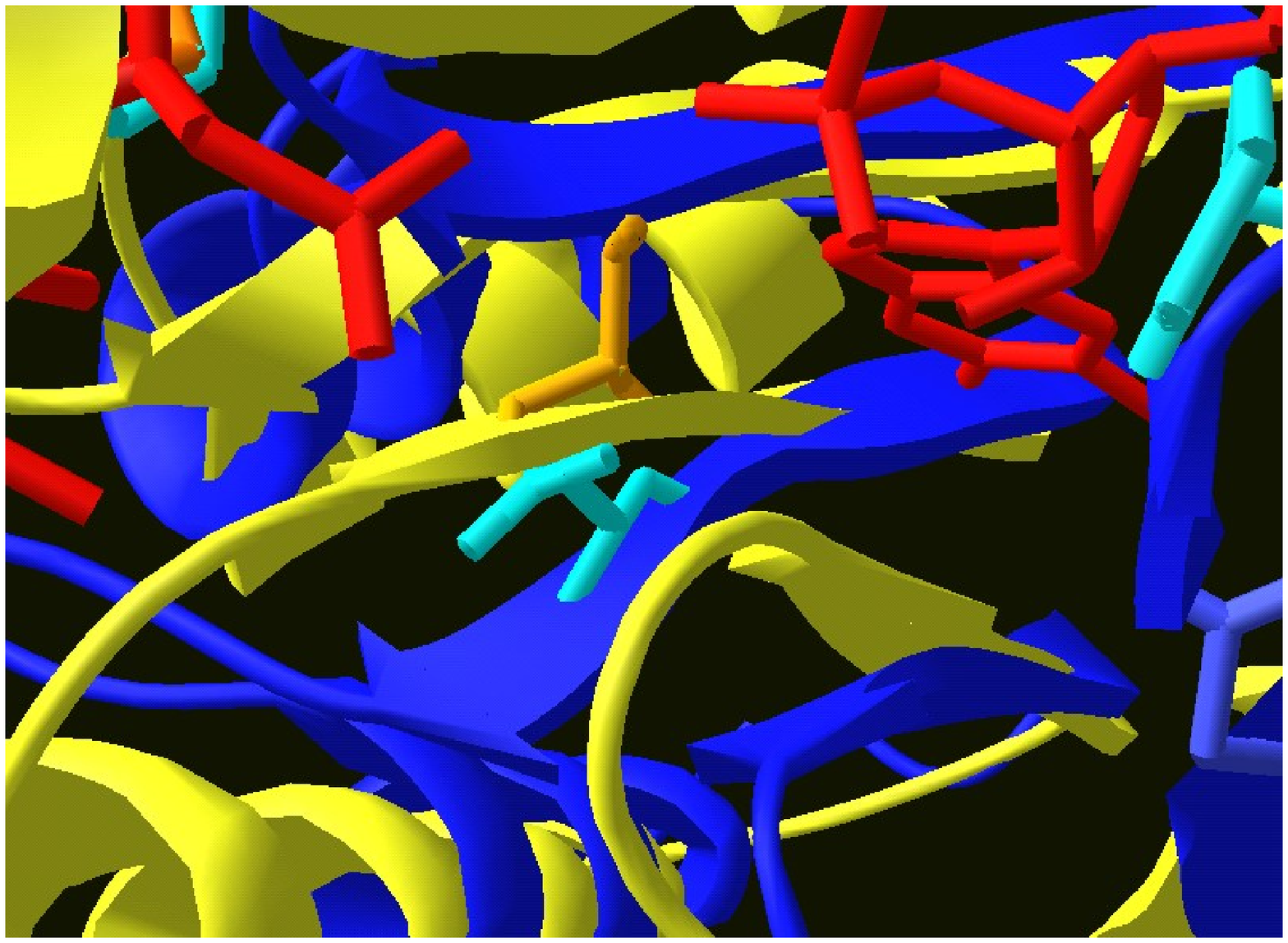}

	{\bf b}
	\includegraphics[width=8cm]{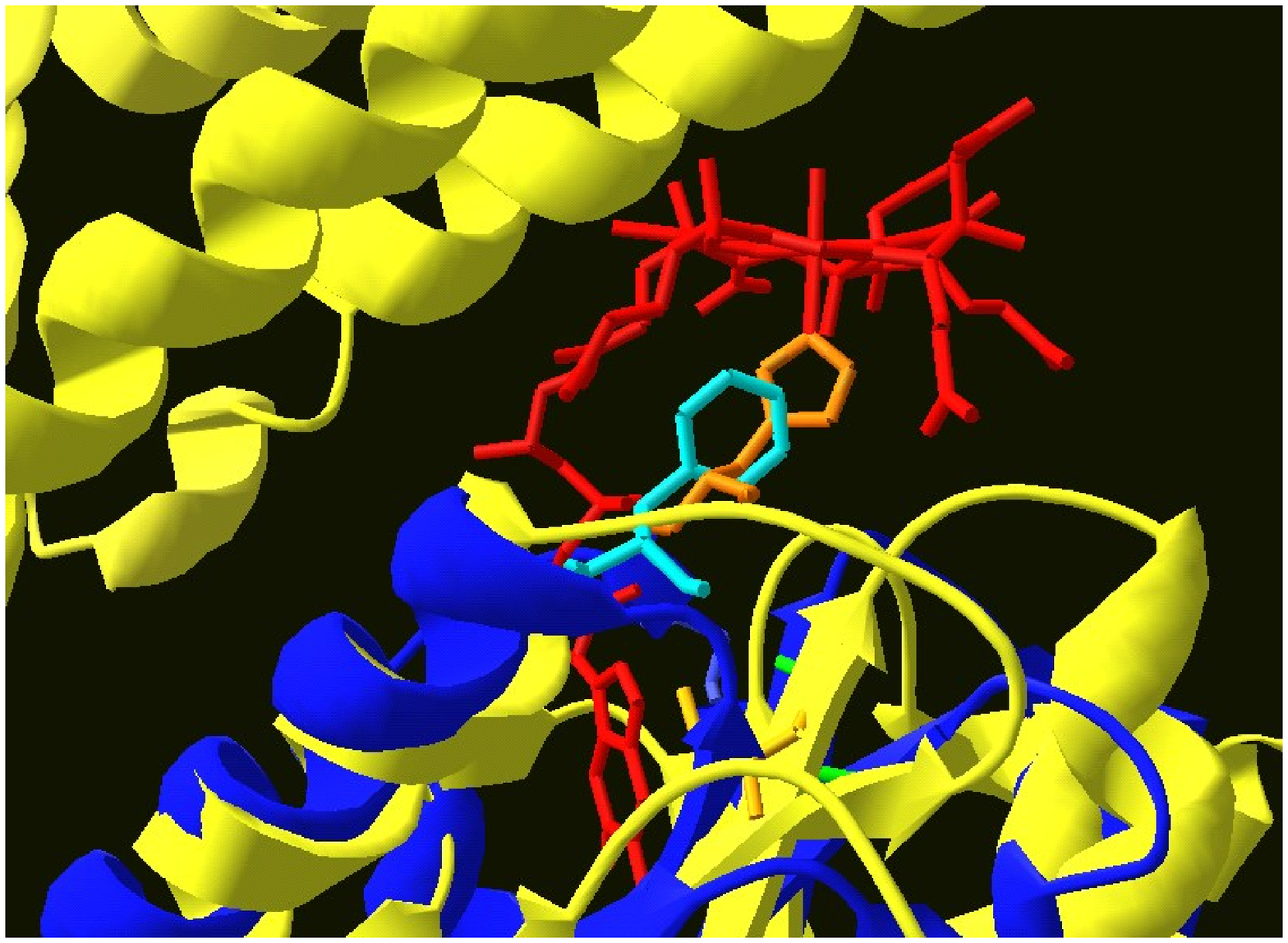}
	\caption{\small Yellow: MetS; blue: CheY; red: cobalamin prosthetic group. 
	         (a) (Center) Orange: MetS:S804; cyan: CheY:D57 (unphosphorylated).
				(b) Orange:MetS:H759; cyan: CheY:F14}
	\label{fig:mets_chey_mpc}
	
\end{center}
\end{figure}

MetS:S804 is an MPC with CheY:D57 (figure \ref{fig:mets_chey_mpc} a).  MetS:S804
hydrogen bonds to N3 of dimethylbenzimidazole, which is the same nitrogen that is
coordinated to the cobalt in free methylcobalamin. The structure of MetS shows this
bond is important for cofactor binding. CheY:D57 is the residue that is
phosphorylated by CheA to activate CheY.

\begin{figure}[h]
\begin{center}
 	\includegraphics[width=10cm]{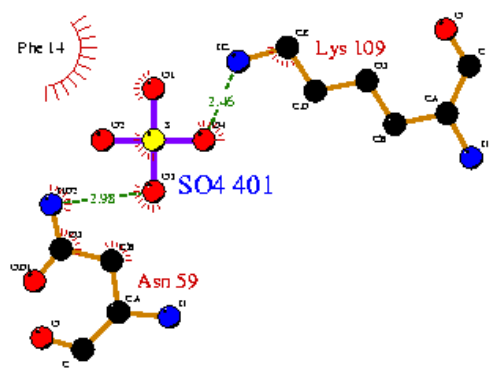}

 	\caption{\small LIGPLOT proposed interaction between the free SO$_4^2-$ and the
	catalytic site in CheY. See text for details.}
	\label{fig:chey_F14_SO4}
\end{center}
\end{figure}

MetS:H759 is an MPC with CheY:F14 (figure \ref{fig:mets_chey_mpc} b).  MetS:H759
axially $\alpha$-coordinates the cobalt atom which lies in the corrin ring. (Axial
$\beta$ coordination is performed from the C-terminal domain, which is not aligned
with CheY, and is therefore not discussed here). CheY:F14 has been the subject of
several mutagenesis studies in Luis Serrano's group
\cite{Bellsolell_etal:96,Wilcock_etal:98}. Serrano \textit{et al.} have shown that
CheY:F14N mutation stabilizes the protein, and accelerates refolding after the
protein has been subjected to urea denaturation. The reason that position 14 is a
capping position of the $\alpha$-helix and 
the asparagine mutant forms a better N-cap.
However, in another study it was shown that the CheY:F14Y
mutation activated CheY constitutively, without phosphorylation
\cite{Bourret_etal:93}.  Although an accurate functional role for CheY:F14 has not
been discovered, it is hypothesized that the proximity to the catalytic residue
CheY:D13 renders it sensitive to mutations, which although serve to stabilize it
(CheY:F14N, and CheY:F14A have shown to lower the unfolding energy of the protein)
probably change the protein's function, as in the CheY:F14Y mutation. A clue as to
CheY:F14's functional role comes from LIGPLOT, a program which automatically
discovers and classifies ligand-protein interactions \cite{Wallace_etal:95}.
CheY:F14 is proposed to form a hydrophobic contact with a free SO${_4}^{2-}$ anion
(figure \ref{fig:chey_F14_SO4}) . This SO${_4}^{2-}$ anion is one of three which
have been identified in the crystal structure. The SO$_4^{2-}$ anion is derived from
the crystallization solution, which contains ammonium sulfate. This particular
anion, is centrally located in the most accessible region of the active site. Its
oxygen atoms are bound to the $\epsilon$-amino nitrogen of Lys-109 and N$\delta$ of
the Asn-59 amide side chain.  As SO$_4^{2-}$ is physically and chemically similar to
PO$_4^{3-}$, it may be representative of how a PO$_4^{3-}$ would interact with the
unactivated CheY in a non-covalent manner. It should be noted that the hydrophobic
interaction CheY:F14--PO$_4^{3-}$, and consequently its supportive role has not been
proposed elsewhere except for by the LIGPLOT diagram. However, taken together with
the constitutive activation of the CheY:F14Y mutation, and the proximity to the
catalytic site, it appears that CheY:F14 does have a functional role having to do
with the binding of the free PO$_4^{3-}$ group.

Table \ref{tab1} shows all MPCs found in the MetS / CheY pair and, for those for which
a structural or functional role has been determined, notes the role. For example,
MetS:P875 and CheY:D38 are both in the N-terminal capping position of their respective
aligned $\alpha$-helices. MetS:T808, a ligand binding residue, is aligned with
CheY:P61 which forms a hydrogen bond with CheY:M63.

Section \ref{sec:case_studies} discusses the implications of structurally aligned
functional residues, in structurally similar, functionally different proteins.

\begin{center}
\begin{table}[hbp]
\small
\label{tab1}
\input{1bmtA03chy00.tab}
\caption{\small {\bf MPCs in MetS and CheY, annotated by function.}
LB -ligand binding; AS-active site member; $\alpha$-C-term: c-terminal acid in an
$\alpha$-helix; $\alpha$-Nc: N-terminal capping residue. Annotations were
determined
according to the crystallographer's papers, the CATH database, and the Protein
Mutant Database (www.genome.ad.jp)}
\normalsize
\end{table}

\end{center}

\chapter{Discussion}
\label{chap:discussion}

This work presents a unique approach to a topic within the framework of
the protein folding problem. It is derived from the premise laid out in the
central dogma of structural biology: ``sequence determines structure''. When
faced with the phenomenon of different sequences which adopt the same fold, a
question which automatically arises is: ``if sequence determines structure,
which elements in different sequences cause them to fold similarly?''

The evaluation of structural / functional importance of discreet amino-acid
positions is not a trivial task.  Indeed, in most cases we cannot know the
role and relative importance of a given residue. Only a handful of proteins
have been investigated thoroughly using site-directed mutagenesis as to the
role, or ``non-role'' as the case may be, of each and every residue.  When
considering a target for site-directed mutagenesis, the positions targeted
are normally purposefully selected, based on prior knowledge of the
protein's sequence to function mapping (or structure to function, if
available). The reason is that setting up a functional assay for most
proteins, and using it to investigate a large amount of mutants is a
laborious process. It is also quite superfluous for most research purposes.
Labor-intensivity and superfluousness hold even stronger when performing
site-directed mutagenesis and setting up an assay to investigate the effects
of mutations on structure, which would require the determination of hundreds
or even thousands of structures for a given protein.

Therefore, it is advisable to turn to computational methods for location
and characterization of critical positions in a protein {\it en-masse}.
Most computational methods base their analysis on evolutionary conservation.
The premise being the following: if a position is determined to be
conserved, then it has been positively selected, and for a good reason.

\section{STAIRS}

Our initial study (Chapter \ref{chap:stairs}) was concerned with Structurally Aligned
Identical ResidueS (STAIRS). The premise being that the few aligned residues
which are identical between proteins differing in sequences in the whole,
are worth investigating as maintainers of structure or function. 
Even better candidates are the STAIRS which are spatially close, which we
named NSTAIRS, (\underline{N}eighboring STAIRS). Those may play a
structural or functional role common to both pair-mates. 

We have determined a conservation score for each aligned position.
Conservation for each position in the database was determined by performing
a multiple alignment using PSI-BLAST. PSI-BLAST was chosen because it
collects and aligns distant family members, in an iterative manner. The
degree of evolutionary conservation was calculated from the last PSI-BLAST
iteration. The reason for that being, that we wanted to look at conservation
which exists in aligned distant family members. This is explained as
follows: conservation of a given position might be due to evolutionary
relatedness of aligned sequences, without actually being important for the
protein's structure.  However, positions conserved between distant sequence
family members are better candidates for being critical positions, as
presumably they are distant enough so that conservation due to evolutionary
non-divergence will be sifted out. Therefore we assessed conservation based
on a sequence alignment of distantly related sequences. 

Normalized conservation scores ($Z_{ic}$ scores, see Chapter
\ref{chap:stairs}/Methods for details) are well correlated between STAIRS and even
better between NSTAIRS.  This initial finding indicated that examining STAIRS and
NSTAIRS as candidates for critical positions is worthwhile. 

However, aligned residues may be identical by chance. 39.5\% of the STAIRS are not
highly conserved. We examined the abundance of STAIRS and NSTAIRS in all the
aligned positions, and in aligned positions which are well conserved
($Z_{ic}\ge1.65$, Chap. \ref{chap:stairs} / Table 1). STAIRS made up 48.4\% of the
population of conserved aligned positions, and 75\% of the well-conserved STAIRS
are NSTAIRS.  This finding set the foundation for the research we conducted and is
elaborated upon in Chapter \ref{chap:mpc}. Namely, look for mutual conservation,
rather than identity \& conservation.

Solvent accessibility is a good initial index for examining location within
the protein. We found that the proportions of buried STAIRS was the same as
that of the entire aligned residue population (\~{}50\%). However, when
examining only conserved STAIRS, the percentage of buried STAIRS was
raised dramatically: \~{}85\%.  We have also partitioned the examined
positions into hydrophobic (HP) and hydrophilic (HY) residue types. In the
entire population, HP residues were more buried than HY types. However, when
looking at the  well-conserved populations, the differences in the ratio of
buried residues partitioned either by hydrophobicity, or according to their
STAIRS/NSTAIRS association were not significant. {\it Circa} 85\% of the
residues were buried.  This is actually expected, as conserved residues,
regardless of physico-chemical traits, are overwhelmingly buried
\cite{Cordes_etal:96}.

In this preliminary study we have shown that positions with conserved, identical
residues may be explained by burial, participation in secondary structure, and by
specific roles pertaining to the functional site in which they reside.

\section{Mutually, Persistently Conserved Positions (MPCs)}

A central observation made in the STAIRS study, was that over 50\% of the
positions with mutually high conservation contain different residues in the
two proteins. This has led us to an extended study, in which we aimed to
characterize the mutually conserved residues in SSSD protein pairs,
regardless of identity. 

\subsection{Substitution matrices}

Our motivations for extracting a substitution matrix from our data were: (1)
analysis of the allowed substitutions in MPC positions; (2) comparison of the
amino-acid distribution of substitutions to the distributions from which the BLOSUM
series has been derived. As shall be explained later, we aimed
to determine the difference between substitutions occurring due to evolutionary
conservation and critical role, and those conserved due to critical role only.

Specialized substitution matrices have been constructed by several research groups,
in order to increase detection sensitivity and alignment reliability, and to study
the substitutions within the data-set. For example, several studies have been
published concerning the substitutions in transmembrane regions of proteins. The
motivation for those studies was that the ``generic'' substitution matrices are
mostly derived from sequences of globular proteins.  Since transmembrane regions of
a protein are in a distinctly different environment, a specialized substitution
matrix would seem more appropriate when analyzing those regions
\cite{Persson_Argos:94,Jones_etal:94,Ng_etal:00}. Indeed, the reported detection
and correct alignment ability achieved by using those matrices for transmembrane
proteins supersedes those of the ``generic'' matrices.  In studies more relevant to
this one, substitution matrices were derived from aligned structures and studied
\cite{Prlic.Domingues.ea:00,Naor.Fischer.ea:96}.  The substitution matrix we have
derived from all aligned residues in our database is very similar to the one
derived by Prlic {\it et al} (2000).

The MPC-derived matrix exhibits some very interesting traits. The most
striking is the high relative entropy (1.015 bits). This relative entropy is
comparable to that of BLOSUM85, which is a matrix derived from BLOCKS
composed of sequences with at most 85\% identity. The high relative
entropy in BLOSUM85 is due to a high rate of synonymous substitutions, which
in turn is due to the fact that the BLOCKS fraction from which BLOSUM85 was
constructed is composed of closely related sequences, which by definition
have a high rate of synonymous substitutions.

The sequences we used to generate the MPC matrix have a very low identity
(\~{}12\%), but the mutually conserved residues we collected had a
high rate of synonymous substitutions. In BLOSUM85, the high rate of
synonymity is due to evolutionary non-divergence among the composing
sequences.  In the MPC-derived matrix, that is evidently not the case. The
hypothesis offered is that the high rate of synonymity here is
due to the irreplaceability of the residues whose substitutions make up the
matrix.

This hypothesis is strengthened by comparing the distances between the
distributions making up the MPC-matrix, and the BLOSUM matrices (Chapter
\ref{chap:mpc}, figure 4). The distribution of substitutions making up the
MPC-matrix differs significantly from those making up the BLOSUM matrices. In
comparison, the distribution of substitutions making up the structurally derived
matrix (SDM) is quite similar to that of BLOSUM35. The latter finding is not
surprising, when we consider that BLOSUM35 is derived from BLOCKS of sequence
alignments with a sequence identity of no more than 35\%. The data-sets from which
both BLOSUM35 and SDM are derived are aligned sequences exhibiting a low
similarity.

An interesting observation regarding the MPC matrix, which was not addressed in
Chapter \ref{chap:mpc}, is offered here.  The score of Ile/Ile replacement is actually lower
than the Ile/Val replacement score (MPC[I,I] $<$ MPC[I,V])
\footnote{MPC[{\em i,j}]
denotes the log-odds value for the substitution of any amino acid {\em i} by
amino-acid {\em j} in the MPC-derived matrix.},
suggesting that replacement of Ile
by Val or vice versa, is superior to Ile's conservation.  The difference between
isoleucine and valine rests with the removal  of a single methyl group substituting
isoleucine's C$\beta$ atom.  It has been shown that Ile$\rightarrow$ Val mutations
lead to an increase in protein stability, up to 2 kcal M$^{-1}$, due to the
deletion of the methyl group \cite{Ventura_etal:02}. It may be hypothesized that
the reason the non-synonymous Ile/Val substitution scores higher than the Ile/Ile
substitution is because of that. On the other hand, stability is not the only
selective factor in a protein's fitness. Another is foldability: $\phi_{\ddag-U}$,
which is measured as the change of stability in the protein's transition state
introduced by a mutation, divided by the change in the stability of the mutated
protein's folded state ($\Delta \Delta G_{\ddag-U}/\Delta \Delta G_{F-U}$).
Negative $\phi_{\ddag-U}$ values can result if the introduced mutation stabilizes
the folded state, but destabilizes the transition state. The Ile/Val mutants
discussed in Ventura {\it et al.} have negative $\phi_{\ddag-U}$ values that may be
the result of strained interactions.  In any case, as the hypothesis that MPC[I,I]
$<$ MPC[I,V] is due to this phenomenon has not been tested, it is presented simply
as an interesting supposal.

\subsection{Cluster analysis}

When predicting the location of critical residues, evolutionary conservation is
seldom the sole criterion for determining the importance of a given position in a
protein. Methods concerned with prediction of critical positions use also
structural information in order to: (1) sift and supplement information given by
evolutionary conservation; (2) analyze conserved positions, in order to determine
{\em why} they are conserved.  In this study, we have also examined the clustering
of MPCs. The term {\it cluster} is used here in a very specific sense: it is a
measureble trait of contacting residues. The quality measured is the inverse of the
probability that those residues shall be in contact, given the sequence distance of
contacting residues in the protein's fold.  In other words, the less probable it is
those residues are in contact, the higher their clustering score.  However, as our
study's main concern was discovery rather than prediction, we used the information
from structural positioning in order to classify the various roles MPCs may play in
defining protein structure.

Spatial proximity of positions provides good augmentative information to
evolutionary conservation. Poteete's studies of critical positions in the T4L
have shown that the critical residues were more
evolutionary conserved than the other positions in the protein
\cite{Rennell.Bouvier.ea:91}. This was shown to be especially true for positions in
the catalytic site, and in those hypothesized to maintain the hydrophobic core. In
both cases those are positions which are both conserved and spatially close.

Shakhnovich and Mirny (1999) have published a study where they examined positions
which are conserved according to the CoC method, and correlated them  with known
folding nucleation centers. Those residues were usually in contact, although they
were not necessarily close in sequence. 

Kannan {\it et al.} have developed a clustering-analysis method using a
weighted-graph representation of the C$\beta$ atoms. The graph is
represented as a Laplacian matrix, and clustering information is derived
from the weighted components of the second lowest eigenvalue. In the first
study of this series \cite{Kannan_Vishveshwara:99} it was observed that
many of the clusters were hydrophobic and buried. However, clusters near
the active and binding sites were also detected. A similar study was
conducted specifically on $\alpha/\beta$-barrel proteins
\cite{Kannan_etal:01_1}, where it was shown that clustered residues are
often conserved, and predicted to be part of the folding nucleus.
Certain clusters were found to be part of the active site, or close to the
active site.

Plaxco \textit{et al.} (1998) have coined the term \textbf{relative contact
order} to describe the average sequence distance between all pairs of
contacting residues normalized by the total sequence length. For all
contacting residues indexed $i,j$:

\begin{equation}
CO = \frac{1}{L\cdot N}\sum^N {\Delta S_{i,j}}
\end{equation}

Where $\Delta S_{i,j}$ is the sequence separation, $N$ is the total number
of contacts, and $L$ is the total number of residues in the protein.  CO is
used to quantify the mean sequence separation between contacting residues in
a chain. In that study, Plaxco \textit{et al.} have shown that relative
contact order is inversely correlated with the folding rate. However, the utility of
contact order may be expanded to examine the clustering of chosen positions
along the structure. The question asked would be: ``given a number of
contacting positions, are they expected to be contacting given their
sequence separation?'' positions which have a low mean sequence separation
are also expected to be in contact. However, widely separated positions
which are in contact, might indicate a functional site, or a structural
stabilizer.

We have developed a novel method for assessing residue clustering. The
method's input is a protein structure, and an indication of the residues
to be analyzed. The method reports whether the indicated
residues are more clustered than expected for the particular protein being
analyzed. A modified version enables us to input two structurally aligned
proteins, and to assess the clustering of indicated residues, MPCs in our
case.

The method is described in detail in the Materials and Methods section in
Chapter \ref{chap:mpc}. Here I will describe its benefits:

\begin{enumerate}

\item {\sc fold specificity}. The method uses a weighted graph to describe
the clustered residues.  The weights of the edges are based upon the
probability of two residues (graph nodes) being in contact ($d(C_A\beta -
C_B\beta)< 7$ \AA), {\em for the given protein}. We propose this weighting
scheme to be better than weighting according to sequence distance, as for
different folds there are different probabilities of two residues being in
the same sequence distance to be in contact.

\item {\sc assessment of statistical significance}. Given a score, we assess
it by a Monte-Carlo procedure. The method is repeated for the protein with
randomly selected positions (or, in the case of MPCs, randomly selected {\em
aligned} positions). In this manner, a distribution of clustering scores
for the number of contacting residues is generated. This distribution is
generated each time the assessment is performed, so that it too is specific
for the protein's (or aligned pair's) structure.

\end{enumerate}

Our clustering assessment method uses the structure of the protein to score
and assess the distance between residues, normalized by the probability of
those positions being in contact. Although we use this method in context
with MPC evaluation, it may be used for any type of clustering evaluation
of selected positions, within a single protein or within a structural
alignment.

Conceivably, this method may be enhanced or modified for a different purpose using
a statistical energy function for contacting residues.  The edge weight for two
contacting residues will be parameterized not only by their sequence distance, but
also by a type-dependent and distance-dependent energy function.  Thus, contact
between residues is not a ``qualitative event'' (i.e. they are either ``in
contact'' or ``not in contact''). Rather, it is a quantitative event.  For example,
suppose that in a given protein two contacting residues, Glu and Arg are separated
by $N$ positions along the sequence, and two other contacting residues, Ala and Ile
are separated by $N$ residues along the sequence.  The graph edge designating the
Glu-Arg contact should be weighted higher than the Ala-Ile contact, as Glu and Arg
may form a salt bridge, whereas Ala-Ile will form a hydrophobic, or a VDW bond.  An
extra refinement could involve the addition of the actual spatial distance between
the residues (based on C$\beta$ atoms, or on centroids), to determine whether
either of the pairs are within an optimal distance for their respective
interactions. To summarize, the introduction of an energy function can contribute
by refining the weight of the graph's edge based not only on sequence distance, but
also on contacting residue type \& spatial distance.


Our clustering-analysis method is a relatively simple one, and derives its
rationale from Plaxco {\it et al.}'s contact order.  However, the
computational overhead is relatively large due to the Monte-Carlo method of
assessment of statistical significance. It is, however a very precise
method, as it checks for statistical significance on a case-by-case basis.

In order to improve the time efficiency of the significance analysis, an analytical
method may be developed based on the analysis of residue clustering distributions
throughout all protein structures. 
The clustering score distribution's parameters can be established empirically in the
following manner:

\begin{enumerate}

	\item Select a group of representative folds (e.g from SCOP).

	\item Establish a distribution of clustering scores for 3,4,5,...
	contacting residues in each fold. The result would be a collection of
	clustering score distributions, parameterized by fold type and number of
	residues in cluster.

	\item When assessing the significance of an MPC clustering score in a
	given SSSD protein pair, the clustering score of the MPCs should be
	compared with that of the appropriate score from the ``distribution
	bank'' above. Given a score $S_{MPC,4}$ for 4 MPCs in a TIM-barrel, the
	score would be compared with the random distribution of scores for 4
	residues in a TIM-barrel fold, in order to determine its significance.

\end{enumerate}

\subsection{Secondary structure analysis}

Another level of determining the structural role of examined residues is to
assess their distribution within secondary structure elements (SSEs).
The distribution of residue types within $\alpha$-helices has been studied
extensively by surveying the residue distribution in helices
e.g. \cite{Richardson.Richardson:88,Aurora.Rose:98,Kumar.Bansal:98}.

In the study presented in Chapter \ref{chap:mpc}, we have looked at the
distribution of MPCs within SSEs in our database. We have aligned the edge and
flanking positions of all SSEs ($\alpha$-helices and $\beta$-strands) of a minimal
length to each other, and looked at flanking and in-SSE positions for high
distributions of MPCs.  Chapter \ref{chap:mpc}, Figure 5 describes our results. In
$\alpha$-helices we found that MPCs are highly abundant in certain flanking
positions. Examining the positions where MPCs were found to be abundant has
revealed that amino-terminal flanking MPCs were mostly residues characterized as
hydrogen-bond acceptors, whereas those flanking the C-terminal were found to be
hydrogen-bond donors. We therefore suggest that in this structural context, MPCs
are important for determining the helix ends by hydrogen bonding to the backbone
atoms.

The residue type distributions in the $\alpha$-helix termini taken from the entire
database were found to be in agreement with previous studies of amino-acid
distributions \cite{Richardson.Richardson:88}. The predominance
of MPCs in the flanking regions, and the over-representation of hydrogen-bond
acceptors
and donors in the N- and C-flanks respectively was quite interesting. To the best
of our knowledge no such systematic investigation of conserved residues along helix
positions has been carried out. The CKAAPS \cite{Reddy.Li.ea:01} study has
determined the frequency of CKAAPS within $\alpha$-helices, but did not
systematically report their relative frequency along the helix itself on a whole
scale database analysis, but only for selected test-cases.

\subsection{Discussion of Case Studies}
\label{sec:case_studies}

Two case studies are presented in, in Chapter \ref{chap:mpc} and in
Chapter \ref{chap:add_res}, and are discussed in detail here.
In chapter \ref{chap:mpc}, a cluster of MPCs from haloalkane dehalogenase from
\textit{X. autotrophicus} (XADL) and lipase B from \textit{C. antarctica} (CALB)
was analyzed. It was shown that those clustered MPCs participated in the active
site, the backbone scaffolding adjacent to the active site, and in substrate
stabilization. XADL and CALB use the same reaction mechanism to catalyze their
reactions, although on different substrates, and with different nucleophiles
(XADL:D124/CALB:S105 ). The other MPC position in this cluster which does not have
the same residue is XADL:W125/CALB:Q106. Possibly this has to do with substrate
specificity, as XADL:W125 is known to play a critical role in XADL's substrate
binding. The other MPCs in this cluster were all glycines, and from the examination
of the structures it appears that they serve as a crucial part of the scaffolding
for the respective proteins. It is interesting to see that mutual conservation can
be used to locate a low common functional denominator, in this case between
functionally related proteins.

In chapter \ref{chap:add_res}, a different pair of proteins is analyzed.
Functionally, MetS and
CheY do not have anything in common. However, several functional
residues are conserved as MPCs. Naturally, it is expected that given a similar
fold, residues that maintain structural similarity will be MPCs, as they should
align within the proteins respective cores, or stabilizing the secondary
structures. Those that were found are described in Table 4.1. Discovering
\textit{functionally} maintained residues as MPCs raises interesting biological and
bioinformatical possibilities which merit further investigation. Biologically, how
common is the phenomenon of functional positions being aligned in proteins which
share the same structure but not the same function? If indeed it were found to be
common, could it be utilized by structural prediction techniques in order to
perform functional predictions? This could be done as follows:

\begin{enumerate}

\item Perform homology modelling, using a template with a known function.

\item Infer from the alignment which residues in the protein with the unknown
function may be responsible for function, based on the functional residues in the
template protein.

\item Use a knowledge based system i.e. a databank of functional sites with an
accompanying search algorithm in order to characterize the function of the protein
with the unknown function. For example, PROCAT  provides facilities for
interrogating a database of 3D enzyme active site templates. PROCAT can be thought
of as the 3D equivalent of the 1D templates found in sequence motif databases such
as PROSITE and PRINTS. Instead of searching for 1D sequence motifs in a newly
derived protein sequence, the PROCAT database allows searching for 3D enzyme active
site template motifs in a protein structure \cite{Wallace_etal:97}.

\end{enumerate}

Currently, many studies are concerned with fold prediction, but another predictive
challenge is that of function prediction, as enzymatic
mechanism determination does not automatically follow from fold prediction
\cite{Erlandsen_etal:00,Todd_etal:99}. 
Locating those residues
suspected of participating in the active sites, and intelligently predicting a
protein's function based on the residues' identities, can
save much of the effort involved with the structural determination of proteins
normally necessary for the elucidation of mechanism of action. Function prediction
can also aid in offering worthwhile alternative hypotheses to current accepted
wisdom. A good example is the alternative mechanism offered by Rigden {\it et al.}
(2000) to the putative nucleotide binding site (NBS) of the R-type plant resistance
genes. Rigden {\it et al.} have suggested that the NBS domain, which was determined
to be so based on sequence motifs only, is actually a phosphorelay domain. This
hypothesis was based on threading, sequence analysis and the construction of a
molecular model. Although no compelling biochemical evidence has been found to
support the phosphorelay hypothesis, it is still accepted as a viable alternative
to the NBS hypothesis \cite{Rigden_etal:00,Fluhr:01,Gebhardt_Valkonen:01}. The MPC
analysis presented here may set the foundations for such a predictive scheme.


\section{Evaluation of PSI-BLAST Alignment Accuracy}

PSI-BLAST was developed as a tool extending BLAST to search for distant
homologues. PSI-BLAST uses a PSSM generated from any given iteration to
extend a search initiated with either a query sequence, or a PSSM generated
by other means. PSI-BLAST has risen to a position of prominence among
bioinformatics tools, and it is probably the most popular tool for searching
for homologues. PSI-BLAST can also be used for fold assignment, by
discovering similarity to the query sequence or the subsequent PSSMs in the
PDB.

PSI-BLAST's ability in fold recognition has been assessed by others
\cite{Park.Karplus.ea:98}.  However, given a true fold assignment, how good is the
actual residue position assignment? The answer to this question is important, as
even for a true fold assignment, a bad alignment can mislead homology modeling.

Using the SSSD database as a benchmark, we have assessed PSI-BLAST's alignment
accuracy of distant homologues. The term ``alignment accuracy'' is a rather vague
one, and requires clarification. Assuming we have a gold standard by which we
assess the alignment accuracy, how is the questioned alignment to be compared with
the gold standard? If the questioned alignment is exactly the same as the gold
standard, then we may assign a top score (e.g. 100\% accuracy). However, in case
the two alignments are not in concord, two separate methods of accuracy assessment
exist:

(1) Sensitivity:

\begin{equation}
	sens = \frac{N_{q\bigcap s}}{N_s}\times 100
\end{equation}

(2) Specificity:

\begin{equation}
	spec = \frac{N_{q\bigcap s}}{N_q}\times 100
\end{equation}

Where $N_q$ is the number of aligned positions in the questioned alignment,
$N_s$ is the number of aligned positions in the gold standard alignment.
$N_{q\bigcap s}$ is the number of aligned positions which exist both in the
gold standard alignment and in the questioned alignment (the size of the
intersecting group between q and s).

So the question ``how accurate is a given alignment method?'' is forked into
two different questions: (1) ``what fraction of truly aligned
positions can a method detect out of the total length of the standard
alignment?'' or ``how sensitive is the alignment method?'' (2) ``what
fraction of the aligned positions in a method's provided alignment are truly
aligned?'' or ``how specific is the alignment method?''

Therefore, no single score can be provided as to the ``accuracy'' of an
alignment method. Rather, when stating alignment accuracy, one must specify
in terms of the method's sensitivity or specificity. In our study, we have
found a significant improvement of alignment sensitivity over consecutive
PSI-BLAST iterations, past the detection iteration. However, no significant
improvement or negative effect was exhibited with regard to specificity. On
the one hand, this means that the number of correctly aligned positions does
increase over consecutive iterations.  On the other hand, the ratio between
correctly aligned positions and incorrectly aligned positions remains
roughly the same. This observation has interesting implications, both
technical and biological. One implication concerns the proper use of
PSI-BLAST: given a query sequence, and an interesting target which the user
would like to investigate further, it is advisable to continue iterating
PSI-BLAST, thus obtaining a more sensitive alignment, one which incorporates
more well-aligned positions. Another implication concerns the proportional
growth of misaligned positions with that of well-aligned positions. This
follows from the observation that over consecutive iterations the
sensitivity increases, but the specificity remains the same. Thus in
absolute numbers, there are more well-aligned positions as the iterations
progress, but in direct proportion to more misaligned positions.
Conceivably, specificity may be improved, by the removal of irrelevant
sequences from the PSSM for the next PSI-BLAST iteration.  This is normally
performed manually, as the choice of sequences to be removed must be made
based on sequence annotation and biological knowledge. Automated improvement
of homology detections in PSI-BLAST, using the literature associated with a
sequence have been performed, although alignment accuracy has not been
reported \cite{Chang_etal:01}.

The biological implication may be stated in the following manner: given two
proteins whose sequence similarity is undetectable by pairwise alignment, detection
by sequence-only considerations is still occasionally feasible.  Given enough
evolutionary sequence-based information (in the form of PSSMs), this information
may be used to obtain alignments as good as those provided by structure prediction
programs using structure-based information, such as threaders.  Our study's
findings correlate well with the results of another study published concurrently
with ours \cite{Sauder_etal:00}.  In that study, structural alignments were derived
using Combinatorial Extension (CE, Shindyalov \& Bourne, 1998) for all superfamily
and family-level related proteins in the SCOP database. PSI-BLAST showed a mean
sensitivity of 40\% after four iterations.

Indeed, this feature of PSI-BLAST has been investigated in the CASP3 and
CASP4 meetings, using several algorithms that were based upon PSI-BLAST
\cite{Moult_etal:01}. One study in CASP3 used a PSSM generated by the query
after four PSI-BLAST iterations through the nr database. This PSSM was used
to search PDB for a suitable homologue, or for fold recognition as the case
may apply \cite{DunbrackRL:99}. As the number of queries evaluated was very
small, the study did not supply a mean alignment accuracy score. Also, some
of the queries had sequence similarity with PDB entries, and some did not,
so that the evaluation of alignment quality for those two different
categories should be done differently. However, alignment specificity
(sensitivity was not given) was shown to be \~{}60\% for those queries with
a low sequence identity with the PDB entries. We have shown an alignment
specificity of $54.9\pm 2.1\%$ in our study.

Another study \cite{Koretke_etal:01} has evaluated the use of a method relying
heavily on PSI-BLAST for fold recognition, SENSER. SENSER attempts to predict a
fold by initially collecting  low e-value hits from a PSI-BLAST run of the query
sequence. (The e-value is described in \ref{sec:psi_blast}).
In this case low e-value hits are sequences with low, but significant
similarity to the query sequence ($< 25\%$ sequence identity, according to Koretke
\textit{et al.})

In the next step, the search is expanded using sequences from the low e-value
collection.  Essentially, a transitive search is performed using low percent
identity, but statistically significant similar sequences. Sequences are aligned
using HMMer (http://hmmer.wustl.edu/), a Hidden-Markov-Model based alignment
method. A study conducted on 15 remote homologues yielded an alignment accuracy of
45\% \cite{Koretke_etal:02}. In this study, alignment sensitivity only is
described.

Another concern is the standard by which alignment accuracy is assessed.
Typically, when assessing an alignment method, the standard referred to is the
structural alignment. However, structural alignments themselves vary
\cite{Godzik:96}, as discussed in \ref{sec:struct_hom}. Therefore, the gold
standard may vary, depending on which structural alignment algorithm is used.  As
stated in the Introduction, we chose the DALI alignment, and verified that it
indeed is in good agreement with SSAP. 

\section{Conclusions}
\subsection{Critical Residues}
When I have started my doctoral studies, the field of critical residue
location by bioinformatical sequence and structure consideration was very
much in its infancy. So was the study of SSSD proteins. In the past three
years we have seen a sharp rise in the number and diversity of papers
discussing both topics. This is due to several reasons:

\begin{enumerate}

\item A steep rise in the number of structures available for
study. Additionally,  a very steep rise in the number of sequences available
for PSSM collation, thus creating more informative PSSMs, commented upon in
Koretke \textit{et al.}, (2001).

\item With the increase in the number of solved structures, SSSD proteins
have been shown to be a common phenomenon. 

\item Sequence-based fold prediction methods have taken a hold in fold prediction,
especially with the introduction of PSI-BLAST, but other methods (e.g.
intermediate sequence search) \cite{Li_etal:00} as well. Taken together with item
(2), there has been a growing interest in the prediction of protein function based
on remote homology. Thus, creating good alignments, rather than just predicting
the fold, has become an important issue.

\item The availability of structural alignment programs, in a usable
manner, opened up this field (among others) to people other than the
authors of those programs.

\end{enumerate}

With the advent of structural genomics, I am convinced that many interesting
structural, functional, and evolutionary connections between different sequence
families populating the same fold will be revealed. There is also an applicable
value for knowing the location and role of critical residues, for protein
engineering and design. For these reasons, the detection and annotation of critical
residues is important, and there is a need for the construction of sensitive
computational methods to do so.

\subsection{Alignment accuracy}

The use of sequence-based predictors will play a major role in the computational
augmentation of the information collected by the crystallographers. PSI-BLAST, by
most accounts, will continue to dominate this area for a while. Therefore, it is
necessary to know how well PSI-BLAST performs in remote homologue alignments, and
what steps should be taken in order to obtain a good alignment. We have studied
this problem using our own restrictive database. Others have offered similar
analyses \cite{Koretke_etal:01,Sauder_etal:00}, although we have managed to
establish the flux in sensitivity and specificity over consecutive iterations. This
information can prove to be very useful, both for remote-based homology modeling,
and as a general guideline for the use of PSI-BLAST, complementing those already
published, (e.g. Jones \& Swindells, 2002).



%% file: 1bmtA03chy00.tab
\begin{tabular}[p]{|l|l|l|l|}
\hline
MetS&role&CheY&role \\
\hline
V750&	&	 L9& \\
H759&AS& F14&LB \\
V766&LB& V21& \\
L770&	&L25& \\
N774&	& G29&$\alpha$-C-term \\
Y775 & &		 F30& \\
I777& & V33& \\
L780&	&	 A36& \\
P785&$\alpha$-Nc&D38&$\alpha$-Nc \\
I789& &	 A42& \\
A798& &	 Y51& \\
D799& &G52& \\
I801&	& V54&  \\
G802& LB&	 I55& \\
S804&AS& D57&AS \\
T808&LB& P61&3-1 Hbond with M63 \\
M821&	& I72& \\
P829&	&P82& \\
L831&LB & L84& \\
G833&LB& V86&G833/G834 are conserved for steric reasons \\
G834&LB& T87& \\
Y850&	&G102& \\
Y856 & &	 Y106& LB \\
N859&LB& P110&Has a cis bond with K109 (active site) \\
A860&LB& F111& \\
\hline
%
%
%
%
\end{tabular}